\documentclass[9pt,conference]{IEEEtran}

\usepackage[dvips]{graphics}
\usepackage[dvips]{graphicx}
\usepackage{amsmath,epsfig,epsf,psfrag,amssymb,amsfonts,latexsym}
\usepackage{verbatim}
\usepackage[mathscr]{eucal}

\def\psfancypar#1#2{\begingroup\def\par{\endgraf\endgroup\lineskiplimit=0pt}
               \setbox2=\hbox{\large\sc #2}
               \newdimen\tmpht \tmpht \ht2 \advance\tmpht by \baselineskip
               \font\hhuge=Times-Bold at \tmpht
               \setbox1=\hbox{{\hhuge #1}}
               \count7=\tmpht \count8=\ht1
               \divide\count8 by 1000 \divide\count7 by \count8 
               \tmpht=.001\tmpht\multiply\tmpht by \count7 
               \font\hhuge=Times-Bold at \tmpht
               \setbox1=\hbox{{\hhuge #1}}
               \noindent
                \hangindent1.05\wd1
               \hangafter=-2 {\hskip-\hangindent
               \lower1\ht1\hbox{\raise1.0\ht2\copy1}%
                \kern-0\wd1}\copy2\lineskiplimit=-1000pt}

\newcommand{\E}{\mbox{{\rm E}}}

\def\boxit#1{\vbox{\hrule\hbox{\vrule\kern3pt
        \vbox{\kern3pt#1\kern3pt}\kern3pt\vrule}\hrule}}

\def\reals{ { {\rm  I \kern-0.15em R }  } }
\def\complex{ {\,{{\rm C} \kern-0.50em \raise0.20ex {  |}}\, }}

\def\Sigmabf{\hbox{$\bf \Sigma$}}

\def\sbf{{\bf s}}

\def\wbf{{\bf w}}

\def\ybf{{\bf y}}

\def\ybf{{\bf y}}
\def\Abf{{\bf A}}
\def\Bbf{{\bf B}}
\def\Cbf{{\bf C}}

\def\Ibf{{\bf I}}

\def\Kbf{{\bf K}}

\def\Pbf{{\bf P}}
\def\Qbf{{\bf Q}}
\def\Rbf{{\bf R}}

\def\Nc{{\cal N}}

\def\Sc{{\cal S}}

\def\Xc{{\cal X}}

\def\be{\vskip .3cm \begin{equation}}
\def\ee{\end{equation} \vskip .4cm \noindent}
\def\defeq{{\stackrel{\Delta}{=}}}
\newcommand{\R}{\mbox{$\hat {\bf R}_{N}$}}

\def\Rxx{\Rbf_{\ssstyle X\kern-.1em X}}

\let\ssstyle=\scriptscriptstyle

\def\Kout{\setbox1=\hbox{\Huge\bf K}\hbox to
1.05\wd1{\hspace{.05\wd1}
}}
\def\Sout{\setbox1=\hbox{\Huge\bf S}\hbox to 1.05\wd1{\hspace{.05\wd1}
}}

  \ifx\LabelFigloaded\MYundefined\relax
  \else
   \fi

  \def\LabelFigloaded{\relax}

  \chardef\LabelFigCatAt\the\catcode`\@
  \catcode`\@=11

 \let\LabelFigwlog@ld\wlog
 \def\wlog#1{\relax}

 \ifx\\\MYundefined@
    \let\\\relax
 \fi

  \def\ms@g{\immediate\write16}

 \def\N@wif{\csname newif\endcsname }
 \def\Temp@ {\N@wif\ifIN@}
 \ifx\INN@\MYundefined@
    \else \let\Temp@\relax
 \fi
 \Temp@

  \def\IN@{\expandafter\INN@\expandafter}
  \long\def\INN@0#1@#2@{\long\def\NI@##1#1##2##3\ENDNI@
    {\ifx\m@rker##2\IN@false\else\IN@true\fi}%
     \expandafter\NI@#2@@#1\m@rker\ENDNI@}
  \def\m@rker{\m@@rker}
 
  \newtoks\Initialtoks@  \newtoks\Terminaltoks@
  \def\SPLIT@{\expandafter\SPLITT@\expandafter}
  \def\SPLITT@0#1@#2@{\def\TTILPS@##1#1##2@{%
     \Initialtoks@{##1}\Terminaltoks@{##2}}\expandafter\TTILPS@#2@}

 \def\Shifted@@#1#2#3{\setbox0=\hbox{#3}%
   \raise -\dp0\vbox {\kern-#2%
       \hbox {\kern#1\unhbox0\kern-#1}%
           \kern#2}}

 \newcount\gridcount
 \newbox\auxGridbox@ \newbox\hGridbox@ \newbox\vGridbox@
 \newbox\Labelbox@ \newbox\auxLabelbox@
 \newbox\Coordinatebox@
 \newtoks\Labeltoks@
 \newdimen\Wdd@ \newdimen\Htt@
 \newdimen\Wddd@ \newdimen\Httt@
 
 \def\Wr@{\immediate\write16}

 \newdimen\GL@wd
 \def\GridLineWidth#1{\GL@wd=#1}

 \def\gobble#1{}
 \def\EdgeErr@{\Wr@{}%
      \Wr@{\string\Edges\space argument
      1, 10, 100 or 1000 please\string!}%
      }

 \newcount\Edgect@

 \def\Sweepup#1\endSweepup{}

 \def\SetEdges@{%
    \edef\Zr@@s{\expandafter\gobble\number\Edgect@\empty}%
        \count255=0\Zr@@s\relax
        \ifnum\count255=\z@\else\EdgeErr@\show\tailtest\fi
        \count255=1\Zr@@s\relax
        \ifnum\count255=\Edgect@\relax\else\EdgeErr@\show\leadtest\fi
    \EdgGl@b\edef\Zr@s{\expandafter\gobble\Zr@@s\empty}
    \ifnum\Edgect@>\@ne\relax\EdgGl@b\let\L@Dc\empty
        \else\EdgGl@b\edef\L@Dc{\string.}\fi
    \ifnum\Edgect@>\@ne\relax
        \EdgGl@b\edef\Edgescale@##1{\divide##1 by \Edgect@}%
        \else\EdgGl@b\edef\Edgescale@##1{}\fi
    }

 \def\Edges#1{\Edgect@=#1\relax
     \let\EdgGl@b\global \SetEdges@}

 \Edges{1}

 \def\hhrule{\hrule height \GL@wd\vskip-.\GL@wd}

 \def\hRule@{%
   \advance\gridcount -2%
   \vfil\hhrule\vfil
   \llap{\smash{\raise -2.5pt
     \hbox{\L@Dc\number\gridcount\Zr@s\kern2pt}}}%
   \hhrule
   }

\def\vvrule{\vrule width \GL@wd \kern-\GL@wd}

 \def\vRule@{\advance\gridcount 2%
   \hfil\vvrule\hfil
   \setbox\auxGridbox@=\vbox to 0pt
      {\vskip \Htt@\vskip 2pt
        \hbox to 0pt{\hss\L@Dc\number\gridcount\Zr@s\hss}\vss}%
      \wd\auxGridbox@=0pt \box\auxGridbox@
   \vvrule
   }

 \def\PlaceGrid@@{\gridcount=10 
  \setbox\hGridbox@=\hbox{%
        \hbox{%
             \hskip-.4pt\vrule
             \vbox to \Htt@{%
               \offinterlineskip\parindent=\z@\relax
               \hbox to \Wdd@{\hfil}
               \hRule@\hRule@\hRule@\hRule@
               \vfil\hhrule\vfil}%
             \vrule\hskip-.4pt}
    }%
  \gridcount=0%
  \setbox\vGridbox@=\hbox{%
      \vbox{\offinterlineskip\parindent=0pt\hsize=0pt
         \vskip-.4pt\hrule%
         \hbox to \Wdd@{%
                 \vtop to \Htt@{\vfil}%
                 \vRule@\vRule@\vRule@\vRule@
                 \hfil\vvrule\hfil}%
         \hrule\vskip-.4pt}}%
  \wd\hGridbox@=0pt\ht\hGridbox@=0pt
  \wd\vGridbox@=0pt\ht\vGridbox@=0pt
  \hbox{\box\hGridbox@\box\vGridbox@}%
  }

 \def\LabelsGlobal{\def\LabGl@b{\global}}
 \def\LabelsLocal{\def\LabGl@b{}}
 \LabelsGlobal 

 \def\SetLabels#1\endSetLabels{%
   \LabGl@b\Labeltoks@={#1()\\}%
   }

 \LabGl@b\Labeltoks@={()\\}

 \def\ShowGrid{\LabGl@b\let\PlaceGrid@\PlaceGrid@@}
 \def\HideGrid{\LabGl@b\let\PlaceGrid@\relax}
 \def\Grids{\ShowGrid\LabGl@b\let\GridSwitch@\ShowGrid}
 \def\noGrids{\HideGrid\LabGl@b\let\GridSwitch@\HideGrid}

 \noGrids

 \def\bAdjust@@{%
     \setbox\auxLabelbox@=\hbox{\raise \dp\auxLabelbox@
            \box\auxLabelbox@}}
 \def\bAdjust@{\let\vAdjust@\bAdjust@@}

 \def\eAdjust@@{\dimen0=-.5\ht\auxLabelbox@
     \advance\dimen0 by .5\dp\auxLabelbox@
     \setbox\auxLabelbox@=
            \hbox{\raise\dimen0\box\auxLabelbox@}}
 \def\eAdjust@{\let\vAdjust@\eAdjust@@}

 \def\tAdjust@@{%
     \setbox\auxLabelbox@=\hbox{\raise-\ht\auxLabelbox@
            \box\auxLabelbox@}}
 \def\tAdjust@{\let\vAdjust@\tAdjust@@}

 \let\vAdjust@\relax

 \def\lAdjust@{\let\hAdjust@\rlap}
 \def\rAdjust@{\let\hAdjust@\llap}

 \let\hAdjust@\relax\let\vAdjust@\relax

 \def\FetchLabel@#1(#2)#3\\{%
     \IN@0#2@@\ifIN@
        \setbox0=\hbox{\ignorespaces#1#3\unskip}%
        \ifdim\wd0>0pt
           \ms@g{}%
           \ms@g{ !!! Bad label(s)? !!!}%
           \message{ #1(#2)#3}%
        \fi
        \def\LabelMole@##1\endFetchLabel@{%
            \IN@0()\\@##1@%
            \ifIN@\def\Temp@{\FetchLabel@##1\endFetchLabel@}%
            \else\def\Temp@{}%
            \fi
            \Temp@
           }%
     \else
       \ignorespaces#1\unskip
       \setbox\auxLabelbox@=%
         \hbox to 0pt{\hss\ignorespaces\hAdjust@
          {\ignorespaces#3\unskip}\hss}%
       \vAdjust@
       \let\hAdjust@\relax\let\vAdjust@\relax
       \AugmentLabelBox@@{#2}%
       \ht\Labelbox@=0pt\dp\Labelbox@=0pt
       \let\LabelMole@\FetchLabel@%
     \fi\LabelMole@}

 \newtoks\XYSep@ 
 \def\SetXYSeparator#1{%
     \IN@0#1@@\ifIN@\XYSep@{*}%
     \else
     \XYSep@{#1}%
     \fi
     }

 \SetXYSeparator*

 \def\AugmentLabelBox@@#1{%
     \IN@0\the\XYSep@ @#1@\ifIN@
       \SPLIT@0\the\XYSep@ @#1@%
       \setbox\Labelbox@=\hbox to 0pt{%
         \unhbox\Labelbox@
         \Shifted@@{\the\Initialtoks@\Wddd@}%
         {\the\Terminaltoks@\Httt@}%
         {\box\auxLabelbox@}}%
     \else
         \ms@g{}%
         \ms@g{ !!! Bad insertion point. !!!}%
         \message{ (#1\ this point was rejected.)}%
     \fi
    }

 \def\FetchOption@#1[#2]#3\endFetchOption@{%
    \def\temp{#1}
    \ifx\temp\empty
       \Edgect@=#2\relax
       \let\EdgGl@b\relax
       \SetEdges@
       \Cleaner@#3%
    \fi}

 \def\Cleaner@#1[@]{\Labeltoks@{#1}}
     
 \def\PlaceLabels@@{\mathsurround=0pt
     \def\Cr@{\\}%
     \let\L\lAdjust@\let\R\rAdjust@
     \let\B\bAdjust@\let\E\eAdjust@\let\T\tAdjust@
     \expandafter\FetchOption@\the\Labeltoks@[@]\endFetchOption@
     \Wddd@=\Wdd@ \Edgescale@\Wddd@ 
     \Httt@=\Htt@ \Edgescale@\Httt@
     \expandafter\FetchLabel@\the\Labeltoks@\endFetchLabel@
     \box\Labelbox@
     }%

 \let \PlaceLabels@\PlaceLabels@@

 \def\AffixLabels#1{\setbox\Coordinatebox@=\hbox{#1}%
      \Wdd@=\wd\Coordinatebox@ \Htt@=\ht\Coordinatebox@
      \advance\Htt@ \dp\Coordinatebox@
      \hbox{\copy\Coordinatebox@\kern-\Wdd@ 
           \Shifted@@{0pt}{-\dp\Coordinatebox@}%
           {\PlaceLabels@\PlaceGrid@}%
           \kern\Wdd@}%
      \GridSwitch@ 
      \LabGl@b\Labeltoks@{()\\}%
      }
 
   \let\wlog\LabelFigwlog@ld   
   \catcode`\@=\LabelFigCatAt  

                                By

              Raymond S\'eroul <A18645@FRCCSC21.BITNET>
                                and 
              Laurent Siebenmann <lcs@topo.math.u-psud.fr>
    
              VERSIONS: July 1991, Oct 1991, Jan 1992, July 1992

INTRODUCTION

      This labelling package is intended for TeX users who
rely on non-TeX sources for for their graphics inserts.  It
provides means for adding TeX labels to such inserts with a
minimum of fuss. 

       For most labels, TeX users have in the past found it
reasonably convenient to rely on non-TeX sources. Typical
occasions when an inescapable need for TeX labels seemed to
arise are

 (a) when the graphics program lacks certain exotic or complex
mathematical symbols

 (b) when the very highest typographical quality is wanted for the
labels

 (c) when labels included with the graphics fail to print, 
 and you cannot figure out why (cf. boxedeps.doc).  The labels
 provided by labelfig.tex are 100

       Since this package first appeared, many users, who in the
past scarcely dreamed of using TeX labels, have come to use
nothing but.  So it is now appropriate to add

Intoxication Warning:  TeX labels may be addictive and expensive. 

     If you have a fast preview you may disagree, and even find
that this package provides an agreeable paste-up environment; see
extra applications at end.

     Note to publishers: It is possible and convenient to ultimately
export the TeX labels produced by labelfig.tex to become an integral
part of the EPS file. This is often desired by a publisher who typically
uses an "upmarket" graphics or page layout program, with which the
staff is skilled in perfecting figures.  See Appendix I for
a recipe.

     The authors are grateful to Patrick Ion of Math Reviews for
helpful comments and encouragement.

BASIC INSTRUCTIONS

    After reading in the macro file using

preview or proof your figure with a coordinate grid printed on
top, by typing the following:

    \ShowGrid  
    \AffixLabels{<the graphics insertion>}

Here <the graphics insertion> is what you would type to insert
the graphics object alone without the grid.  This must provide
for the space around it. For example <the graphics insertion>
might well be \BoxedEPSF{MyFigure scaled 700} using the
boxedeps.tex macro package (from same source); this provides a
TeX box containing the encapsulated PostScript insert specified by
the file MyFigure. \AffixLabels{...} provides the grid (supposing
\ShowGrid is present) and later, once you have specified labels
using the grid, it will "tack on" the labels.

     The grid is a sort of (usually elongated) checkerboard of
ten rows and ten columns and its (internal) partitions are by
default numbered  .1, ... ,.9  both horizontally (X-coordinate
running left to right) and vertically (Y-coordinate running bottom
to top).  Thus the points enclosed by the grid correspond to the
points of the unit square in the cartesian "X-Y" plane, the lower
left corner corresponding to the origin (0,0).  By extrapolation,
the full page corresponds to a larger rectangle in the plane.

     These coordinates serve to position labels as follows.
Before the \AffixLabels{...} command type label specifications:

  \SetLabels
   (<X-coordinate>*<Y-coordinate>) <first label> \\
   .
   .
   .
   (<X-coordinate>*<Y-coordinate>)  <last label> \\
  \endSetLabels

Each row specifies one label and is terminated by \\.  In each
row, the position indicator comes first; it is written as a
standard cartesian point except that the X- and Y- coordinates
are separated by * rather than a comma because TeX allows a
comma as decimal point. There are no dimension units to specify
as the unit is the grid itself.

     By default, this cartesian point specifies where the middle
of the baseline of the label will be located.  However if you precede
the point by \L [or \R] the left [or right] edge of the baseline will
be located there. Similarly you may also precede the point by \T, \E,
or \B to vertically align the top equator or bottom of the label box
at the specified point.  This gives nine standard positions of
the label with respect to the insertion point --- corresponding to
the eight principle points of the compas and the center

                     \L\T     \T      \R\T

                     \L\E     \E      \R\E

                     \L\B     \B      \R\B

But this neglects the default "baseline" level of TeX,
giving potentially three more positions

                     \L    <no tag>   \R

For text, the baseline level is often the preferred. Its relation to
the others is variable. It will often coincide with the bottom level,
as happens for "X".  But it is often distinct, as for "g", in which
case you have in all 12 distinct positions rather than 9.

     It is convenient to think of this specification of label
position as attaching the label by a thumb-tack to the coordinate
grid. There are up to twelve positions of the thumb-tack on the
label, while the position of the thumb-tack on the coordinate grid is
arbitrary.  Normally, one choses the position of the thumb-tack on
the label to be the one that is the closest to the item being
labeled.  There are good reasons for this "rule of thumb":

   (a)  It facilitates correct positioning at first try.

   (b)  If the scale of the figure must be altered after labels
have been affixed, the labels have a good chance of remaining well
positioned.

   (c)  The visible grid need not extend beyond the "bounding box"
for the figure, because the best preferred position is always
(at least almost) within the bounding box .

The second reason is particularly important. Indeed it often
happens that scale has to be altered after labelling begins, in
order to either provide space for the labels, or to adjust
proportions between the labels and the figure.  (The size of labels
is unaffected by scaling.)

     Here is an artificial but self-contained test which uses
TeX rules to make a graphics object.

TEST

    Do not skip this!

 \def\FrameIt#1{\hbox{\vrule$\vcenter {\hrule\kern3pt%
             \hbox {\kern3pt #1\kern3pt}%
               \kern3pt\hrule}$\relax\vrule}}

 \def\Caption#1#2{\FrameIt{%
       \vtop {\hsize=#1\relax \parindent=0pt
         \leftskip=0pt \rightskip=0pt plus15pt
         \parfillskip=0pt
         \lineskip=1pt\baselineskip=0pt
         #2}}}

 \def\FirstQuadrant{\hbox to 100pt{\vrule\vbox to 100pt{%
        \hbox to 100pt{\hfil}\vfil\hrule}\hss}}

  \SetLabels
    \R(.5*.2) $\zeta\,\cdot$\\
    (.9*-.10) $\xi$\\
    \R(-.03*.9) $\eta$\\
    \T(.5*.9) \Caption{70pt}{%
          \it The norm of
          $g(\xi+i\eta)$ is indicated on
          contours of this invisible surface.}\\
  \endSetLabels

  \AffixLabels{\FirstQuadrant}

  \end

  Note that the coordinates to use for labels are indicated on the
edges of the grid (when visible) corresponding to the conventional
x- and y- axes of the Cartesian plane. By default the grid is
1-by-1. However, by the command \Edges{100}, you can change this
to 100-by-100 and many users find this alternative most
convenient. Place the command \Edges{...} in your style file (or
header) since its effect is is global. Other possible edge values
are 10 and 1000.

  If you use the command \Edges{...} at all, do so with care.  For
if you accidentally delete an \Edges{...} command your labels will
abruptly be badly misplaced and may logically but mysteriously
generate "dimension too big" errors under TeX and "off page" errors
under your driver.  

  You can dictate the edgescale for an individual figure by giving
the scale in brackets immediately after \SetLabels.  Thus, to
import into an article using say \Edge{100} a figure labelled using
another edgescale, say the original 1-by-1 default, you can use
\SetLabels[1]...\endSetLabels.

GETTING IT DOWN PAT

     Complicated labeling deserves the same respect as
complicated mathematics.  Do not expect it to come out perfect the
first time!  What is needed in either case is a mechanism to
repeatedly typeset troublesome pieces.

     One mechanism is always available.  One does complicated
labelling in a separate "test" file involving just the figure being
labelled;  a texpert will know how to \dump TeX's current state as
a temporary format that restarts rapidly at each retry.  Usually,
one then pastes the completed labelled figure back into the main
TeX file, but, of course, one can also \input it as an auxiliary
file.

     If you do not have a TeXpert at handy, here is a first
approximation to an efficient setup. By deletions reduce a copy
of your article to just a few lines before and after the figure.
Now label the figure, and finally, copy and paste the labelled
figure to the original article. Then copy the next figure to label
into this testbed and repeat. The TeXpert can improve the  speed
at which TeX starts up, by compiling a format specifically for
your article; just one caution: best NOT include in the format
ephemeral details of setup like \Set<mydriver>ArtSpecials (from
boxedeps.tex because this reads  figure dimensions which you may
change during your work session.

     An improved mechanism to repeatedly typeset troublesome
pieces is now available on the Macintosh; it is called LinoTeX;
see the same ftp sources.  It could be set up on many types
of computer.

     Before using labelfig.tex to attach labels to a graphics
object inserted using boxedeps.tex or BoxedArt.tex, make it a
firm rule to carefully adjust the bounding box using the trimming
commands of these packages, and also at least tentatively scale
and position the object. Beware of changing the grid inadvertently
after the labels have been positioned.  For example, correcting
the bounding box of a PostScript graphics object can foul up the
labels by changing the coordinate grid to which the labels are
attached. This is particularly true for the trimming  commands of
boxedeps.tex and BoxedArt.tex. However, as noted already, change
of scale is much less disruptive, and modest adjustments should be
well tolerated.

     Sometimes the labels protrude so far from the bounding box
of a figure that the figure has to be repositioned.  Best do this
by ad hoc spacing, say using \hglue and \vglue; altering the
bounding box would create a vicious circle.

     Remember that you are responsible for preventing labels
from overlapping. You are responsible for all label typography
including size and style. A label is really just about anything
that can be put in a TeX box. Note that spaces at the beginning
and end of labels will normally be suppressed; if you really want
them you must protect them with TeX braces.

     This package temporarily sets the \mathsurround parameter
of TeX to zero  while the labels are being affixed. This is done
because nonzero \mathsurround space would influence the position
of left and right aligned labels; then, when a texpert or printer
modifies mathsurround, diagram labeling might be disastrously
altered. There is a small price to pay involving labels that are
formatted as caption boxes including mathematics: you  may want or
need to specify an explicit mathsurround space within the caption
box; it will not influence anything outside.

     Those hostile to the use of * as separator between
the X and Y coordinates of label insertion points, are free to
impose another using \SetXYSeparator{<the new separator>}.  
Americans may prefer "," to "*" since they never use a 
comma as a decimal point; on the other hand, * may be more visible.

APPENDIX (I)  MERGING labelfig.tex LABELS INTO AN EPSF GRAPHICS OBJECT.

     As promised in the introduction, here is a recipe useful for
publishers. It works at least on Macintosh and at least for vectorized
graphics and Adobe type1 fonts.  (There is surely a similar recipe for
PCs under MSWindows.)

 (a)  Use boxedeps.tex utility to integrate the figure given by the eps
file, "x.eps" say, with a visible frame around it.  See
\ShowDisplacementBoxes command in boxedeps.tex.  To get precise results
automatically it is important to use the \Trim... commands of
boxedeps.tex making the "DisplacementBox" neatly fit the figure.

 (b)  Use the TeX printer driver and LaserWriter (versions >= 8.1.1) to
export to an EPSF the DVI page containing the integrated, labelled
figure. You now have an EPS file  "xx.eps"  that contains too much, and at
the wrong scale, and at wrong position.

 (c)  Convert the EPSF to an Adode Illustrator format EPSF using
the shareware utility called epsConvert by Sam Weiss
1993-- (currently $25).

 (d)  In Illustrator (or a compatible program), group the labels and the
"DisplacementBox"; copy them to the clipboard and paste them into "x.ps".
This step requires that all the label fonts be "visible to the Macintosh.

 (e)  Translate and scale the pasted group consisting of the labels plus
the "DisplacementBox" so as to make the "DisplacementBox" the bounding
box of (labelless) figure represented by "x.eps".  At this point the
labels will be correctly placed on the figure "x.eps".

 (f)  Ungroup and delete the "DisplacementBox".  The result is the
desired single EPS file, "x+.eps" say, It contains the original figure
plus its labels.  

     Using grouping and ungrouping appropriately in "x+.eps", a
publisher's staff can very efficiently improve label positions etc.

APPENDIX II)  SOME EXOTIC APPLICATIONS

     The grid of labelfig.tex is analogous to a light-table in
classical page makeup with wax or latex glue.  In principle, you
can use it to compose any page from its indivisible parts.  This
even has some of the artisanal charm of classical paste-up
provided you have a fast screen preview to make the process
"interactive".

     In practice labelfig.tex is a tool for nonstandard jobs.
Here are a few going beyond the labelling already discussed.

(I)  GRAPHICS INTEGRATION.

     This is accomplished by treating the imported graphics
objects as labels.  The underlying graphics object is then
typically an empty  \vbox to <dimension>{\vfill} in a TeX
\midinsert...\endinsert construction.  A label line
might be of the form

   (.1*.1) \\

The exact form of the special command varies from driver to
driver.  However, in the case of encapsulated PostScript graphics
(EPSF norm), by relying on boxedeps.tex, one can have the
following standard syntax (independant of driver  (see
boxedeps.doc for details.
  
  (.1*.1) \BoxedEPSF{MyFigure scaled <scale in mils>}\\

This may be slow since it requires TeX to read the PostScript
file to read bounding box using many complex macros.  So you
may want to try

  (.1*.1) \EPSFSpecial{MyFigure}{<scale in mils>}\\

which is fast and driver independant, but it squashes the
bounding box, normally to its lower left corner.

     Similarly for graphics of the Macintosh PICT norm ---
using BoxedArt.tex (same sources) in place of boxedeps.tex.

     This approach to integration is to be recommended when
one is assembling a composite graphics object.

 (II)  COMMUTATIVE DIAGRAM ENHANCEMENT

     Commutative diagrams or arrays of mathematical objects
connected by arrows of various sorts are common in mathematics.
The mathematical objects require the use of TeX.  Recently TeX
acquired a good collection of arrows of all slopes --- that of
LamSTeX --- plus pwerful macros to build the diagrams.

     However, even the LamSTeX collection is often
inadequate; it lacks for example double shafted arrows, dotted
arrows and curved arrows. Fortunately it is possible to produce
such arrows on an individual basis using sophisticated graphics
programs such as Illustrator and AldusFreehand (both serving
the EPSF norm) or using Metafont (with its public domain norm).
Since the creation of each new arrow is a work of love, you
probably want to limit the number of arrows by using LamSTeX
for most arrows. The 40K commutative diagram module of LamSTeX
has been adapted to work with AmSTeX and a copy may be posted
with LabelFig and related files. Unfortunately no one has yet
offered a version that works with Plain TeX or LaTeX.

       Suffice it here to say that when the exotic arrow has
been somehow imported into TeX, labelfig.tex treats it as a
label that one affixes to the commutative diagram.  Two other
steps will be treated in separate notes, namely the matter of
extracting the dimension specifications for the arrow and the
construction of the arrow --- for these steps are far from
unique and often depend intimately on your computer environment. 
Notes for the Macintosh-Textures-Illustrator combination are
found in the file ExoticArrows.doc.

 (III) NESTING 

Ingenuity pays off in exploiting labelfig.tex. One can
mix graphics and typography quite freely.  labelfig.tex is good
for freeform or overlapping arrangements, while boxedeps.tex (or
BoxedArt.tex) is best for regimented non-overlapping
arrangements --- and the two can be combined.

     The default behavior of labelfig.tex is not ideal 
for nesting objects, because to prevent trouble for beginners
the register for labels is globally cleared when \AffixLabels
concludes.  But there are switches available

      \LabelsGlobal      \LabelsLocal

which change this.  To understand this, extend the above test 
by something like:

 \LabelsLocal

 \SetLabels
    (.5*.5) AAA\\
 \endSetLabels

 {
 \SetLabels
    (.5*.5) ZZZ\\
 \endSetLabels
   \AffixLabels{\FirstQuadrant}
 }

   \AffixLabels{\FirstQuadrant}

     There are however potential pitfalls.  Neither
labelfig.tex nor boxedeps.tex has been tested under extreme
conditions. Problems may occur if their procedures are
indiscriminately nested. For boxedeps.tex (not labelfig.tex)
there is a precise cause for worry, namely many of its
variables are "global", which means that TeX braces will not
provide the protection one might expect.

COMMAND SUMMARY FOR labelfig.tex

  Here [...] means optional (one or zero)
       [...]* means any number of such constructs

  \SetLabels
    [[<P>](<X><Sep><Y>) <label> \\]*
  \endSetLabels
  \ShowGrid  
  \AffixLabels{<the figure>}

   --- <P> is tack position, one of eleven or empty
              order irrelevant

                   \L\T      \T      \R\T

                   \L\E      \E      \R\E

                     \L               \R

                   \L\B      \B      \R\B

   --- (<X><Sep><Y>) insertion point;
  <Sep> is separator, = * by default;
  \SetXYSeparator{<Sep>} changes it.
   <X> and <Y> are real numbers

  --- <label> a label to attach 

  --- <the figure> the figure to label 

  \GlobalLabels (default)     
  \LocalLabels  setting for nested constructs.

 \Grids makes ALL grids appear; \HideGrid then makes just next disappear.
 \noGrids returns to default.  The commands are always global.

 \GridLineWidth{<dimension>} adjusts width of grid lines. Default is very
small, to give "hairline" effect. If your grid lines are missing try
setting \GridLineWidth{1pt}.

 \Edges#1 globally changes the edge size of all grids to the numerical 
value #1, which must be 1, 10, 100, or 1000.  The default is 1.

VERSION HISTORY.
 --- Jan 1993: \Edges#1 and [??] option after \SetLabels
 --- July 1992: \Grids, \noGrids, \HideGrid;
       Gridlines become hairlines; \GridLineWidth{<dimension>}.
 --- Oct 1991, Jan 1992: \SetXYSeparator{<Sep>},  \LabelsGlobal,
       \LabelsLocal.
 --- July 1991: first release

Address for bugs and other feedback:

        Raymond S\'eroul
        IREM and Lab. de Typographie Informatise
        Univ. Rene Descartes
        Strasbourg

    Tel 33-88-41-63-45
    Email:  A18645@FRCCSC21.BITNET

        Laurent Siebenmann
        Mathematique, Bat. 425,
        Univ de Paris-Sud,
        91405-Orsay,
        France

    Tel 33-1-6941-7949; 
    Email: lcs@topo.math.u-psud.fr

\newtheorem{proposition}{Proposition}

\newtheorem{theorem}{Theorem}

\def\nn{\nonumber}
\def\defeq{\stackrel{\Delta}{=}} 
\def\scalefig#1{\epsfxsize #1\textwidth}
\def\Ebb{{\mathbb E}}

\newcommand{\SNR}{\mbox{SNR}}

\newcommand{\Xmsc}{\mathscr{X}}

\newcommand{\beq}{\begin{equation}}
\newcommand{\eeq}{\end{equation}}

\IEEEoverridecommandlockouts
\title{Sensor Configuration and Activation for Field Detection\\ in Large Sensor Arrays}
\author{\authorblockN{ Youngchul Sung and Lang Tong\thanks{This work was supported in part by
the Multidisciplinary University Research Initiative (MURI)  under
the Office of Naval Research Contract N00014-00-1-0564.  Prepared
through collaborative participation in the Communications and
Networks Consortium sponsored by the U.~S. Army Research
Laboratory under the Collaborative Technology Alliance Program,
Cooperative Agreement DAAD19-01-2-0011.}}
        \authorblockA{School of Electrical and Computer Engineering\\
        Cornell University\\
	Ithaca, NY 14850, USA \\
        Email:\{ys87,ltong\}@ece.cornell.edu}
        \and
        \authorblockN{H. Vincent Poor\thanks{The work of H. V. Poor was
supported in part by the Office of Naval Research under Grant
N00014-03-1-0102.}}
        \authorblockA{ Dept. of Electrical Engineering\\
          Princeton University \\
          Princeton, NJ 08544 \\
          Email:poor@princeton.edu}}

\begin{document}
\maketitle

 {\footnotesize
\begin{abstract} The problems of sensor configuration and activation for
the detection of correlated random fields using large sensor
arrays are considered. Using results that characterize the
large-array performance of sensor networks in this application,
the detection capabilities of different sensor configurations are
analyzed and compared. The dependence of the optimal choice of
configuration on parameters such as sensor signal-to-noise ratio
(SNR), field correlation, etc., is examined, yielding insights
into the most effective choices for sensor selection and activation in various operating regimes.
\end{abstract} }
\vspace{-0.5em}
\section{Introduction}
\label{sec:intro}

\begin{figure}[hbtp]
\centerline{
\begin{psfrags}
\psfrag{A}[c]{Access point} \psfrag{B}[l]{Sensor}
\psfrag{C}[c]{Sensor Network}
\scalefig{0.35}\epsfbox{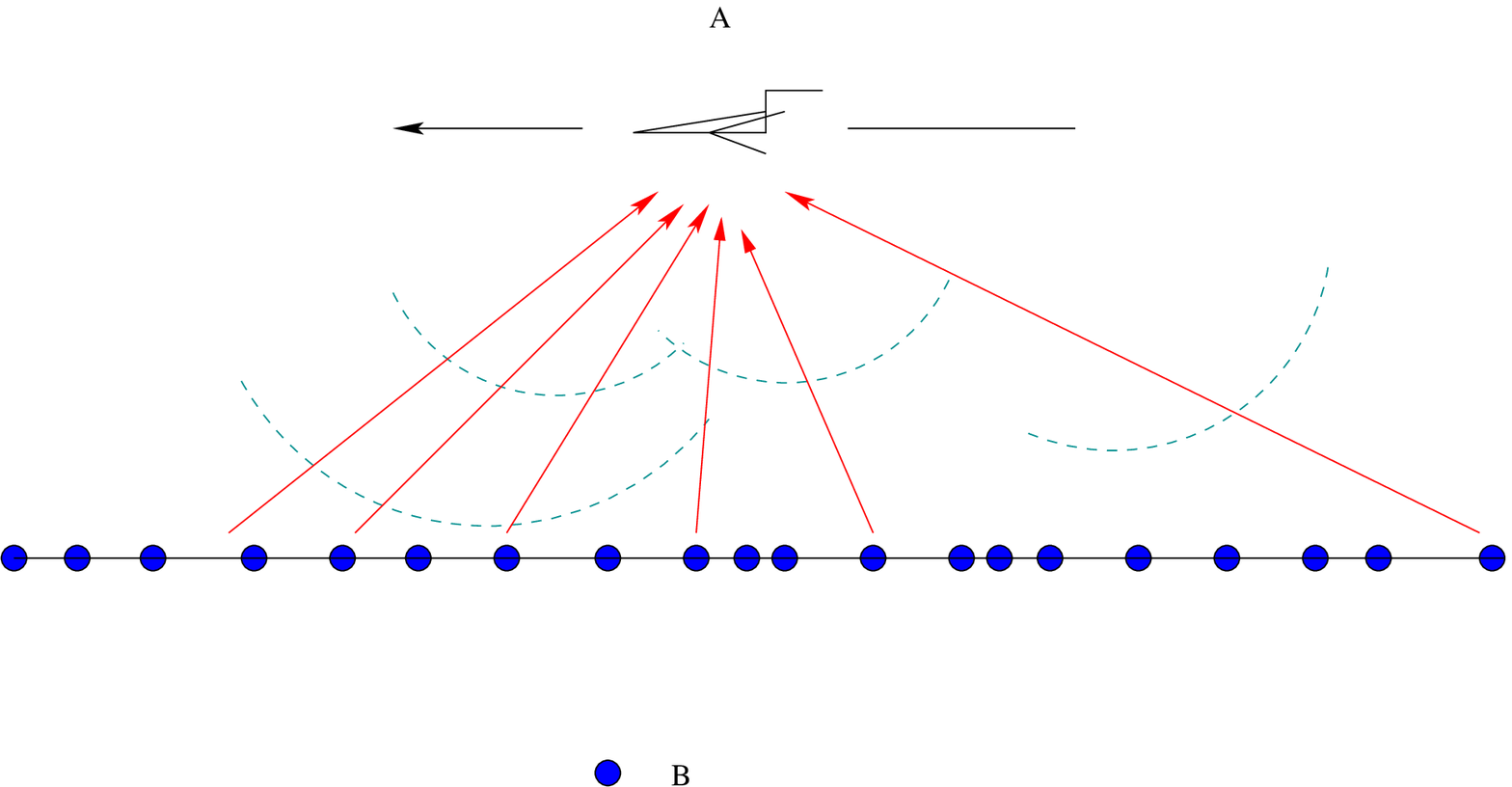}
\end{psfrags}
} \caption{Sensor network with  mobile access point}
\label{fig:senma}
\end{figure}

The main design  criteria for  sensor networks are the performance
in  the specific task and the energy efficiency of the network. In
this paper, we consider optimal sensor configuration and selection 
for densely deployed sensor networks for the detection of
correlated random fields. An example in which the problem of such
sensor selection  arises is Sensor Network with Mobile Access (SENMA), as
shown in Fig.~\ref{fig:senma}, where a mobile access point
collects sensor data controlling sensor transmissions in the
reachback channel.
To maximize the energy efficiency of such a network, one should
judiciously select and activate sensors to satisfy the desired
detection performance with the minimum amount of sensor data since
the number of activated sensors is directly related to the energy
consumption of the entire network.

To simplify the problem for analysis, we focus on a 1-dimensional
space, and investigate how various parameters such as the field
 correlation, signal-to-noise ratio (SNR), etc., affect the optimal
 configuration
 for different sensor schedules.
\begin{comment}
\begin{figure}[htbp]
\centerline{ {
    \begin{psfrags}
    \psfrag{0}[c]{ $0$}
    \psfrag{L}[c]{ $x_n$}
    \psfrag{a}[r]{ $s(x)$}
    \psfrag{d}[c]{ $\Delta_{i}$}
    \psfrag{xi}[c]{ $x_i$}
    \psfrag{xi1}[c]{ $x_{i+1}$}
    \scalefig{0.3}\epsfbox{figures/sample_ra.eps}
    \end{psfrags}
} } \caption{Signal and sensor location} \label{fig:sensorlocations}
\end{figure}
\end{comment}
Specifically, we assume that the signal field $s(x)$ is the
stationary solution of the stochastic diffusion equation \cite{Cox&Miller:book}:
\begin{equation} \label{eq:diffusioneq}
\frac{ds(x)}{dx}= - A s(x)+ B u(x), ~~x\ge 0,
\end{equation}
where  $A \ge 0$ and  $B$ are known, and the initial condition is
given by $s(0) \sim \Nc(0,\Pi_0)$. Here, $x$ denotes position
along the linear axis of the sensor array. The process noise
$u(x)$ is a zero-mean white Gaussian process, independent of
 both sensor measurement noises $\{w_i\}$ and the initial state $s(0)$.
We assume that  each activated sensor takes a
measurement of the field at its location, and subsequently
transmits the data to the collector or fusion center\footnote{
 We will not focus on local quantization of $y_i$ at the sensor
level here, nor will we consider the transmission error to the
fusion center. These are important design issues that must be
treated separately.}. The observation $y_i$ from the activated
sensor $i$ located at $x_i$ ($x_i < x_{i+1}$) is governed by the
following statistical hypotheses
\begin{equation}  \label{eq:hypothesisscalar}
\begin{array}{lcl}
H_0 &: & y_i = w_i, ~~~~i=1,2,\cdots, n,\\
H_1 &: & y_i = s_i+ w_i, \\
\end{array}
\end{equation}
where $\{w_i\}$ are $\Nc(0,\sigma^2)$ measurement noises, with
known $\sigma^2$ and independent from sensor to sensor, and
 the signal sample $s_i \defeq s(x_i)$. The dynamics of the collected signal samples $\{s_i\}$ are given by
\begin{eqnarray}
s_{i+1} &=& a_i s_i +  u_i,  \label{eq:statespacemodelgeneral}\\
 a_i   &=&  e^{-A\Delta_{i}}, \label{eq:statespaceai}
 \end{eqnarray}
 where  $\Delta_{i} \defeq |x_{i+1}-x_i|$
and $u_i \sim \Nc(0, \Pi_0(1-a_i^2))$. A similar model was derived
in \cite{Micheli&Jordan:02MTNS}.

 Note that $0 \le a_i \le 1$ for
$ 0 \le  A \le \infty$ and $a_i$ determines the amount of
correlation between sample $s_i$ and $s_{i+1}$; $a_i=0$ implies
that two samples are independent while for perfectly correlated
signal samples we have $a_i=1$ . By the stationarity, $\Ebb
\{s_i^2\} = \Pi_0$ for all $i$, and the SNR for the observations
is given by ${\Pi_0}/{\sigma^2}$.

\subsection{Summary of Results}

 We adopt the Neyman-Pearson formulation of fixing
the detector size $\alpha$ and minimizing the miss probability.
The miss probability $P_M(\Xmsc,n;\alpha,\mbox{SNR})$ is a
function of the number, $n$, and locations, $\Xmsc
\defeq \{x_1,\cdots,x_n\}$, of the activated sensors as well as detector size
$\alpha$ and SNR. Usually, the miss probability decreases
exponentially as $n$ increases and the error exponent is defined
as the decay rate
\begin{equation}\label{eq:exp}
K_\alpha(\Xmsc;\SNR)=\lim_{n\rightarrow \infty} \frac{1}{n}\log
P_M(\Xmsc,n;\alpha,\mbox{SNR}).
\end{equation}
The error exponent is a good performance index since it gives an
estimate of the number of samples required for a given detection
performance; faster decay rate implies that fewer samples are
needed for a given miss probability. Hence, the energy efficient
configuration for activation  can be formulated to find the optimal $\Xmsc$
(where data should be collected)  maximizing the error exponent
when the sample size is sufficiently large.

 Based on our
previous results on the behavior of the error exponent for the
detection of correlated random
fields \cite{Sung&Tong&Poor:04ITsub}, we examine several strategies
for  sensor configuration for the testing of $H_1$ versus $H_0$, and
propose guidelines for the optimal configuration for different
operating regimes. Specifically, we consider uniform configuration,
periodic clustering, and periodic configuration with arbitrary sensor
locations within a period.  We show that the optimal configuration
is a function of the field correlation and the SNR of the
 observations.  For uniform configuration, the optimal strategy 
is to cover the entire signal field with the activated sensors for
SNR $>1$. For SNR $<1$, on the other hand, there exists an optimal
spacing between the activated sensors. We also derive the error
exponents of  periodic clustering and arbitrary periodic
configurations. Depending on the field correlation and SNR, the
periodic clustering outperforms uniform configuration. Furthermore,
there exists an optimal cluster size for intermediate values of
field correlation. The closed-form error exponent obtained for the
vector state-space model explains the transitory error behavior
for different sensor configurations as the field correlation
changes. It is seen that the optimal periodic configuration is
either periodic clustering or uniform configuration for highly correlated or
almost independent signal fields.

\subsection{Related Work}

The detection of Gauss-Markov processes in Gaussian noise is a
classical problem. See \cite{Kailath&Poor:98IT} and references
therein. Our work is based on the large deviations results in
\cite{Sung&Tong&Poor:04ITsub}, where the closed-form error
exponent was derived for the Neyman-Pearson detection, with a
fixed size, of correlated random fields using the innovations
approach for the log-likelihood ratio (LLR) \cite{Schweppe:65IT}.
There is an extensive literature on the large deviation approach
to the detection of Gauss-Markov
processes \cite{Benitz&Bucklew:90IT}-\cite{
Luschgy:94SJS}.
   The application of the large deviations principle (LDP) to
   sensor networks has been considered by other authors as well. The
sensor configuration problem can  be viewed as a sampling problem. To
this end, Bahr and Bucklew \cite{Bahr&Bucklew:90SP} optimized the
exponent numerically under a Bayesian formulation. For a specific
signal model (low pass signal in colored noise), they showed that
the optimal sampling depends on SNR, which we also show in this
paper in a different setting.  Chamberland and  Veeravalli have
also considered the detection of correlated fields in large sensor
networks under the formulation of LDP and a fixed
threshold for the LLR test with the focus on detection performance
under power constraint \cite{Chamberland&Veeravalli:04ITWS}.

\section{Preliminaries: Error Exponent and Properties}
\label{sec:preliminaries}

In this section we briefly present previous results
\cite{Sung&Tong&Poor:04ITsub} relevant to our sensor configuration
problem.
 The error exponent for the Neyman-Pearson detection of the hypotheses
 (\ref{eq:hypothesisscalar})
with a fixed size $\alpha \in (0,1)$ and uniformly configured
sensors with spacing $\Delta$ (i.e., $\Xmsc = \{
(i-1)\Delta\}_{i=1}^n$) is given by
\begin{equation}
K_\alpha(\Xmsc;\SNR)= -\frac{1}{2} \log\frac{\sigma^2}{R_{e}} +
\frac{1}{2} \frac{\tilde{R}_{e}}{R_{e}}
 - \frac{1}{2}, \label{eq:errorexponentscalar}
\end{equation}
independently of the value of $\alpha$,
 where $R_{e}$
 and $\tilde{R}_{e}$ are the steady-state
  variances of the innovations process of $\{y_i\}$ calculated under $H_1$ and $H_0$, respectively.
The closed-form formula (\ref{eq:errorexponentscalar}) is obtained
via the innovations representation of the log-likelihood ratio
\cite{Schweppe:65IT}, and enables us to further investigate the
properties of the error exponent with respect to (w.r.t.)
parameters such as the correlation strength and SNR. (See
\cite{Sung&Tong&Poor:04ITsub} for more detail.) Here, we note that
the error exponent for the miss probability with a fixed size does
not depend on the value $\alpha$ of the size. Thus, the error
exponent depends only on $\Xmsc$ and SNR. (For notational
convenience, we use $K$ for the error exponent unless the
arguments are needed.)

We now describe the basic properties of the error exponent
starting from the extreme correlation cases. 

\vspace{0.3em}
\begin{theorem}[Extreme correlations] \label{theo:extremecor}
The error exponent $K$ is a continuous function of the correlation
coefficient $a \defeq e^{-A\Delta}$ for a given
 SNR. Furthermore,
\begin{itemize}
\item[(i)] for i.i.d. observations ($a=0$) the error exponent $K$ reduces to the  Kullback-Leibler
information $D(p_0||p_1)$ where $p_0 \sim \Nc(0,\sigma^2)$ and $p_1
\sim \Nc(0, \Pi_0+\sigma^2)$;
\item[(ii)] for the perfectly correlated signal ($a=1$) the error exponent $K$ is zero for any SNR,
and the miss probability decays to zero  with
$\Theta(\frac{1}{\sqrt{n}})$.
\end{itemize}
\end{theorem}

\vspace{0.3em}
The above theorem  reduces to the Stein's lemma
for the i.i.d. case.
 For the perfectly correlated case ($a=1$), on the other hand, the miss probability does not decay exponentially;
 rather it
 decays in polynomial order $n^{-1/2}$.

 The error behavior for intermediate values of correlation is
 summarized by  the following theorem, and shows distinct characteristics for different SNR regimes.

\vspace{0.3em}
\begin{theorem}[$K$ vs. correlation]  \label{theo:etavscorrelation}
\begin{itemize}
\item[(i)] For SNR $> 1$, $K$ decreases monotonically as the
correlation increases (i.e. $a \uparrow 1$); \item[(ii)] For SNR
$< 1$, there exists a non-zero correlation value $a^*$ that
achieves the maximal $K$, and $a^*$ is given by the solution of
the following equation.
\begin{equation}  \label{eq:optimalam}
[1+a^2+\Gamma(1-a^2)]^2-2(r_e+\frac{a^4}{r_e})=0,
\end{equation}
where $r_e = R_{e}/\sigma^2$. Furthermore, $a^*$ converges to one
as SNR approaches zero.
\end{itemize}
\end{theorem}
Hence,  an i.i.d. signal gives the best detection performance for
a given SNR $> 1$.  The intuition behind this result is that the
signal component in the observation is strong at high SNR, and the
new information contained in the observation provides more benefit
to the detector than the noise averaging effect present for
correlated observations.  For SNR $< 1$, on the other hand, the
error exponent does not decrease
 monotonically as correlation becomes strong, and there exists an optimal  correlation.  This is because at
low SNR
 the signal component is weak in the observation and correlation between signal samples
 provides a
 noise averaging effect.  This noise averaging  will become evident in Section \ref{subsec:periodiccluster}.
\begin{figure}[htbp]
\centerline{ \SetLabels
\L(0.25*-0.1) (a) \\
\L(0.75*-0.1) (b) \\
\endSetLabels
\leavevmode
\strut\AffixLabels{
\scalefig{0.23}\epsfbox{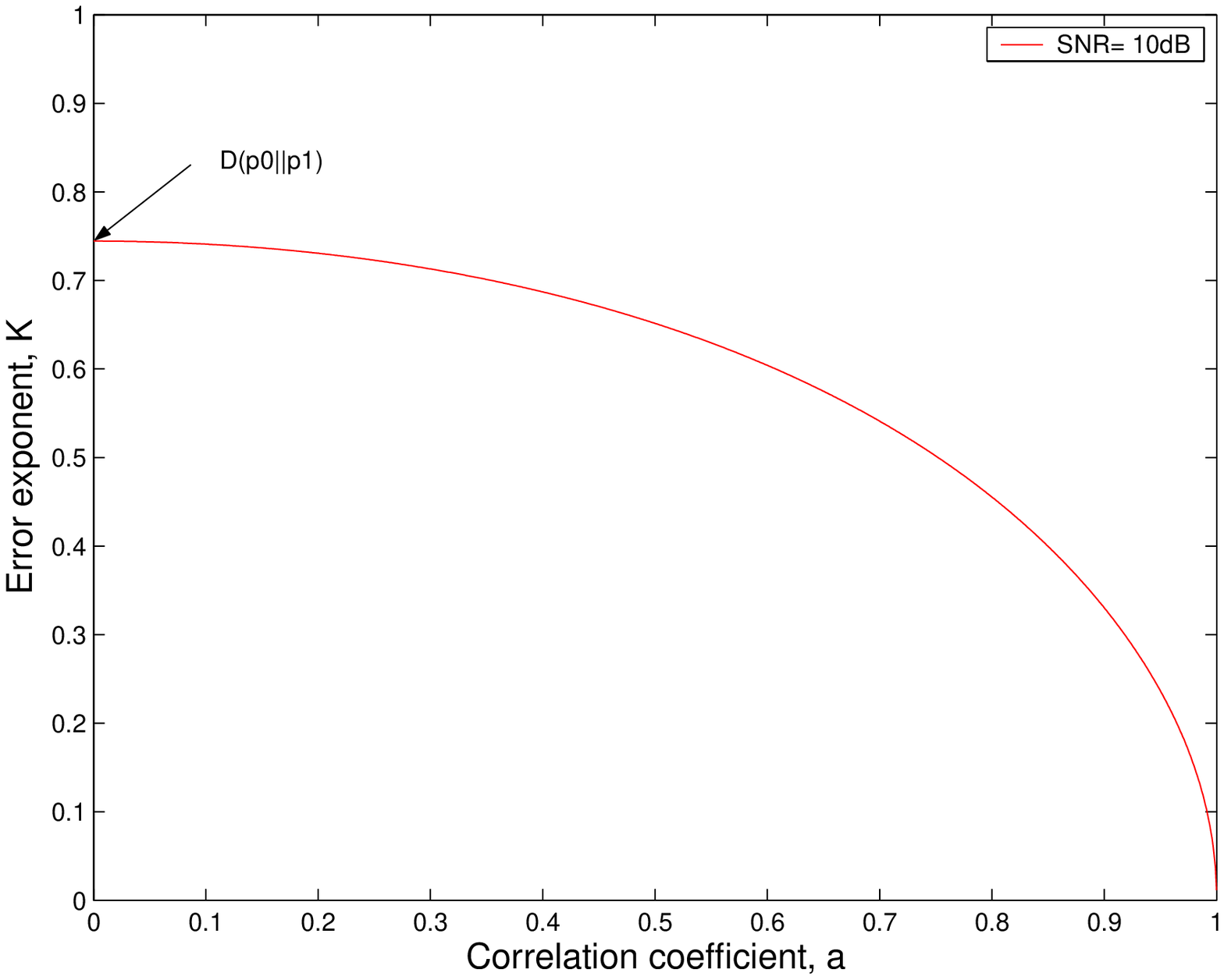}
\scalefig{0.234}\epsfbox{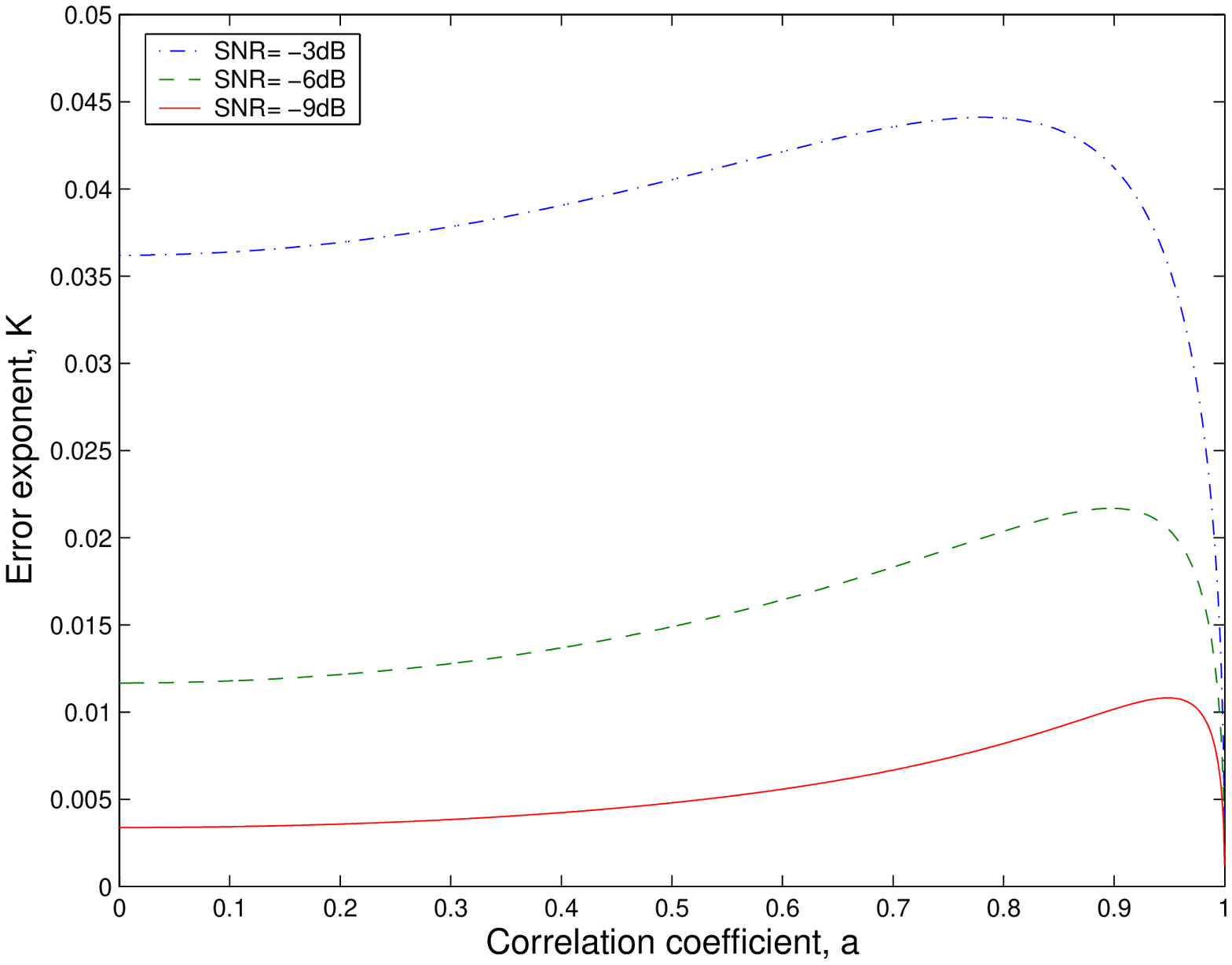} } }
\vspace{0.5cm} \caption{$K$ vs. correlation coefficient $a$: (a)
SNR = 10 dB (b) SNR= -3, -6, -9 dB} \label{fig:Kvsa}
\end{figure}
Fig. \ref{fig:Kvsa} shows the error exponent as a function of the
correlation coefficient
  $a$ for several values of SNR.  Two plots clearly show the different error behaviors as a function
  of  correlation
   in the  high and low SNR regimes.   Unit SNR is a transition point between two different
   behavioral regimes of the error exponent as a function of correlation strength.

The error exponent is also a function of SNR.  This aspect of the
behavior of the error exponent is given by the following theorem.

\vspace{0.3em}
\begin{theorem}[$K$  vs. SNR] \label{theo:etavsSNR}
The error exponent  $K$ is monotone increasing as SNR increases
for a given correlation coefficient $0 \le a <1$.  Moreover, at
high SNR  the error exponent $K$ increases linearly with respect
to $\frac{1}{2}\log \mbox{SNR}$.
\end{theorem}

\section{Sensor Configuration}
\label{sec:optimalscheduling}

\begin{figure}[hbtp]
\centerline{
\begin{psfrags}
\psfrag{a}[c]{(a)} \psfrag{b}[c]{(b)} \psfrag{c}[c]{(c)}
\psfrag{d}[c]{(d)} \psfrag{Delta}[c]{$\Delta$}
\psfrag{MD1}[c]{$M\Delta$} \psfrag{MD2}[c]{$M\Delta$}
\psfrag{Xc}[l]{$\Xc$} \psfrag{sensor}[l]{\textsf{Sensor}}
\scalefig{0.350}\epsfbox{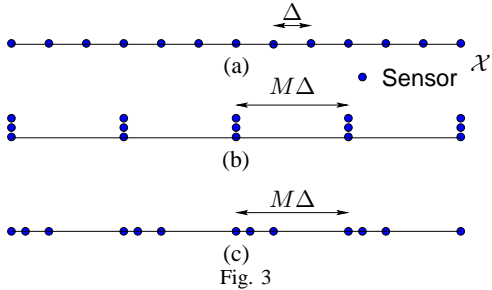}
\end{psfrags}
} \caption{Configurations for $n$ sensor activations:
(a) uniform configuration (b) periodic clustering (c) arbitrary
periodic configuration} \label{fig:schedulingprotocols}
\end{figure}

In this section, we investigate several sensor configurations, and analyze
 the corresponding detection performance via the error exponent.
 We also provide the closed-form error-exponent for several interesting cases by extending the
results in the previous section. Specifically, we consider uniform
configuration, clustering, and periodic configuration with arbitrary
locations within a spatial period as described in Fig.
\ref{fig:schedulingprotocols}. We provide the optimal
configuration for uniform configuration for the detection of
stationary correlated fields, and investigate the benefit of other
configurations.

\vspace{-0.5em}
\subsection{Uniform Configuration} \label{subsec:uniformschedule}

\begin{figure}[hbtp]
\centerline{
\begin{psfrags}
\scalefig{0.30}\epsfbox{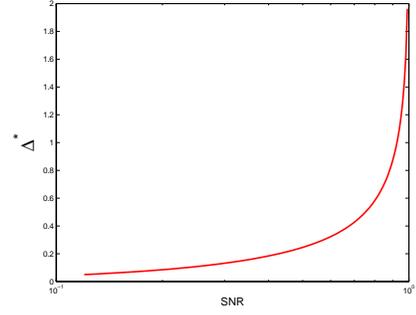}
\end{psfrags}
} \caption{Optimal spacing between sensors for infinite signal
field ( SNR $<1$, $A=1$)} \label{fig:lowSNRoptimaldistance}
\end{figure}

For  uniform configuration with spacing  $\Delta$ between neighboring
sensors, the data model is described by the state-space model
 (\ref{eq:statespacemodelgeneral}) with
 \[
 a_1=\cdots=a_n=a = e^{-A\Delta},
 \]
 and the results in Section \ref{sec:preliminaries} are
directly applicable.  The key connection between sensor configuration
and detector performance is given by the correlation coefficient
$a$. First, we consider $\SNR
> 1$. In this case, by Theorem \ref{theo:etavscorrelation} {\em (i)}, the error exponent decreases
 monotonically as $a$ increases, i.e., the spacing  $\Delta$ decreases for a given field diffusion rate  $A$.
Hence, when the support of the signal field $\Sc$ is finite and
$n$ sensors are planned to be activated in the field,
 the optimal uniform scheme is to distribute the $n$ activated sensors to cover all the signal field, which makes
  the observations least correlated; localizing all the scheduled sensors in a subregion of the stationary signal field
 is not optimal.  For $\SNR < 1$, on the other hand, the optimal spacing $\Delta^*$ for an infinite (in size) signal field
 is given by
 \begin{equation}
\Delta^* = - \log \frac{a^*}{A},
 \end{equation}
where $a^*$ is given by the solution of (\ref{eq:optimalam}).
 $\Delta^*$ is finite for any SNR strictly less than one since the diffusion coefficient
 $A < 0$  and $a^* > 0$ for any SNR $< 1$. The optimal spacing as a function of SNR is shown for $A=1$
in Fig. \ref{fig:lowSNRoptimaldistance}.

\begin{figure}[htbp]
\centerline{ {
    \begin{psfrags}
    \scalefig{0.3}\epsfbox{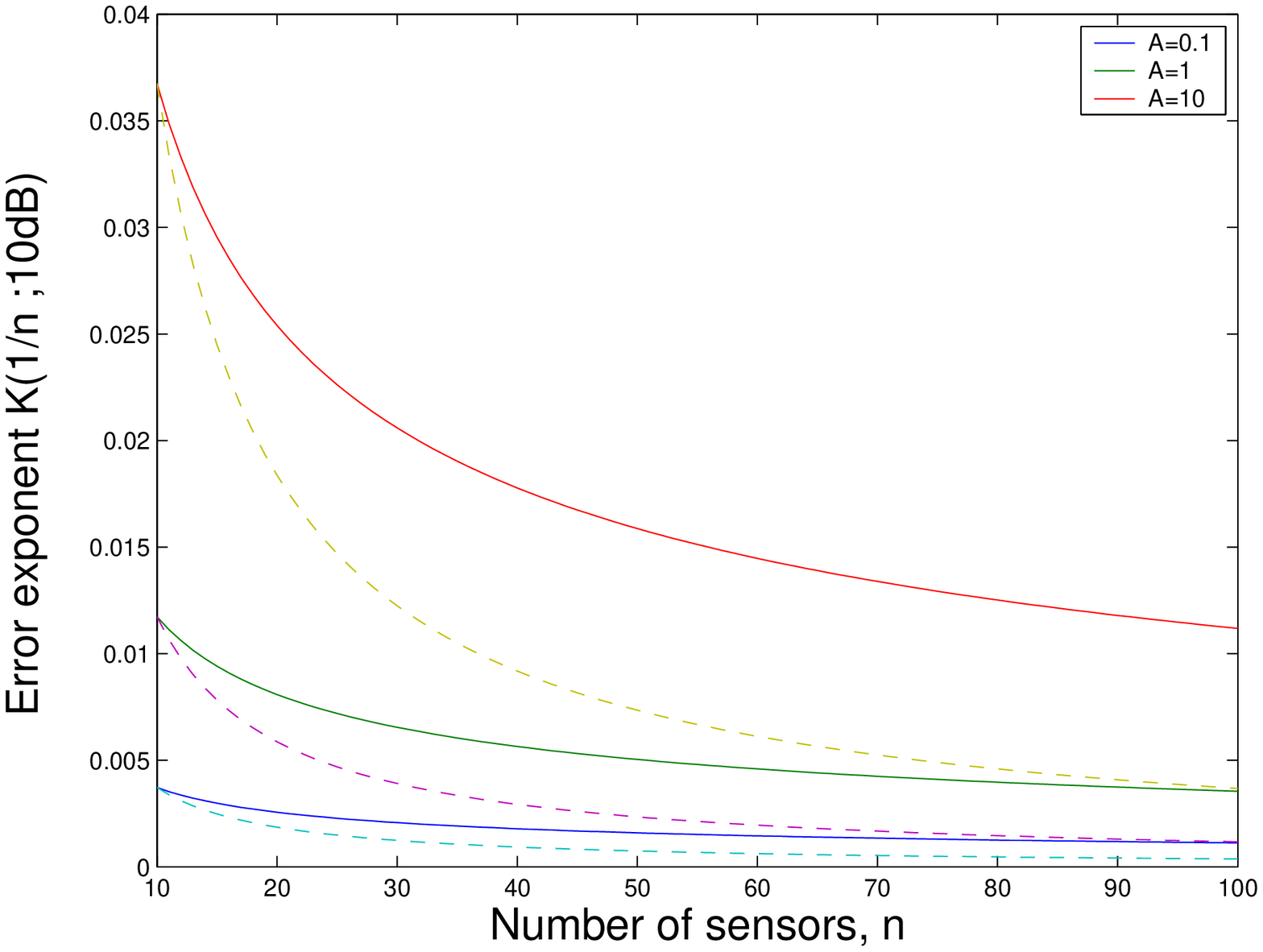}
    \end{psfrags}
} } \caption{Error exponent vs. $n$ for a fixed signal field
($|\Sc|=1$)} \label{fig:finitefieldSNR10dB}
\end{figure}

For a finite signal field $\Sc$ with $n$ scheduled sensors,
$\Delta^*$ is still optimal among the class of uniform
configurations if $n\Delta^* < |\Sc|$, where $|\Sc|$ is the
spatial duration of the signal field. In this case, the sensor
field does not need to cover the entire  signal field. However, if
$n\Delta^*
> |\Sc|$, $\Delta^*$ may no longer be  the optimal spacing.  As shown in Fig.
\ref{fig:Kvsa}, the error exponent decreases when the spacing
$\Delta$ deviates from $\Delta^*$. Hence, activating sensors fewer
than $\bar{n}
\defeq \lfloor \frac{|\Sc|}{\Delta^*}\rfloor$ with spacing larger
than $\Delta^*$ always
 gives a worse performance than $\bar{n}$ sensors with
 spacing $\Delta^*$. However, this may not be the case for activating more
sensors than $\bar{n}$ (up to $n$) by reducing the spacing  from
$\Delta^*$. Even if the error exponent decreases by reducing the
spacing, more sensors are activated over the signal field.
Therefore,  better performance is possible for the latter case
since the product of the error exponent and the number of samples
determines the miss probability approximately. Similar situation
also occurs at SNR $>1$ for  finite signal field. Note that the
error exponent increases as the correlation decreases. (See Fig.
\ref{fig:Kvsa}.) Thus, by spreading the activated sensors with a
reduced number of activated sensors in the signal field, sensor
data become less correlated and the slope of error decay becomes
larger at a cost of reducing the number of observations. However,
the increase in the error exponent is not large enough to
compensate for the loss in the number of sensors in the field.
Fig. \ref{fig:finitefieldSNR10dB} shows the error exponent as a
function of the number of scheduled sensors in $\Sc$ ( $|\Sc|=1$)
at 10 dB SNR for several different diffusion rates. The dashed
line shows the decay of $n^{-1}$ for which the performance loss by
the decrease in the number of scheduled sensors is exactly
balanced. We observe that the decay of the actual error exponent
is slower than $n^{-1}$. Hence, when the maximum number of
available sensors in a finite signal field case is
$n$, the optimal configuration is to activate all $n$ sensors
covering the entire field with maximal spacing.

Another interesting fact about the finite signal field is the
asymptotic behavior  when the number of sensors increases without
bound. In this case, the correlation coefficient converges to one,
i.e.,
\begin{equation}  \label{eq:sensorplaceinfinitydensity}
a = \exp \left(- A \frac{|\Sc|}{n}\right) \rightarrow 1
~~~\mbox{as}~ n \rightarrow \infty.
\end{equation}
By Theorem \ref{theo:extremecor} {\em (ii)}, the error probability
does not decay exponentially, but decays with polynomial order
$n^{-1/2}$ for any finite $A$ as $n\rightarrow \infty$. The
exception is the singular case where $A = \infty$, i.e.,
 the signal is a white process.  Therefore, for the detection of stationary correlated
 fields, it is a better strategy to cover a larger area as long as
 the signal field extends there than to localize
 activated sensors  more densely in a subregion.

\subsection{Periodic Clustering} \label{subsec:periodiccluster}

The uniform configuration for a finite signal field reveals that
there is a benefit at high SNR to making sensor spacing large to
obtain less correlated observations,  but activating fewer sensors
results in a bigger loss than the gain from being less correlated.
This naturally leads to our second configuration:
periodic clustering shown in Fig. \ref{fig:schedulingprotocols}
(b),  aiming at the benefits from both correlation and the number
of scheduled sensors. In this configuration, we activate $M$
sensors very close in location, and repeat this multiple
activation periodically over signal field so that the number of
scheduled sensors is preserved and the spacing between clusters
becomes larger than that of uniform configuration.

For further analysis, we assume that the $M$ sensors within a
cluster are located at the same position.  With the total number
of scheduled sensors $n=MN$, the observation vector $\ybf_n =
[y_1,y_2,\cdots, y_n]^T$ under $H_1$ is given by
\begin{equation}
\ybf_n = \tilde{\sbf}_n \otimes {\mathbf 1}_M  + \wbf_n,
\end{equation}
where $\otimes$ is the Kronecker product,
\begin{equation}
\tilde{\sbf}_n =[s(0), s(\tilde{\Delta}),\cdots,
s(N\tilde{\Delta})]^T,
\end{equation}
and $\tilde{\Delta} = |\Sc|/N= M \Delta$. ($\Delta$ is the sensor
spacing for the uniform configuration for $n$ sensors in $\Sc$.) The
covariance matrix of $\ybf_n$ is given by
\begin{equation}\label{eq:covarianceclustering}
\Ebb \{\ybf_n \ybf_n^T\} = \left\{
\begin{array}{cc}
\Sigma_{s,N}(\tilde{a})  \otimes {\mathbf 1}_M{\mathbf 1}_M^T  +\sigma^2\Ibf& \mbox{under}~H_1,\\
\sigma^2 \Ibf & \mbox{under}~H_0,
\end{array}
\right.
\end{equation}
where $\tilde{a}= \exp(-A \tilde{\Delta})$.
 The signal covariance matrix has a
 block Toeplitz structure due to the perfect correlation of
signal samples  within a cluster.  $ \Sigmabf_{s,N}(\tilde{a})$ in
(\ref{eq:covarianceclustering})
 is a positive-semidefinite Toeplitz matrix where
the $k$th off-diagonal entries are given by $r_s(k)=\Pi_0
\tilde{a}^k$. For any $A > 0$, ~$0 \le \tilde{a} < 1$ and
$r_s(\cdot)$ is an absolutely summable sequence; the eigenvalues
of $\Sigmabf_{s,n}$ are bounded from above and
below\cite{Gray:72IT}.  Using the convergence  of the eigenvalues
of $\Sigmabf_{s,N}$ and the properties of the Kronecker product,
we obtain the error exponent for periodic clustering.

\begin{proposition}[Periodic Clustering] \label{prop:K_period_clustering}
For the Neyman-Pearson detector for the hypotheses
(\ref{eq:hypothesisscalar}) with level $\alpha \in (0,1)$ and
periodically clustered sensor configuration, the error exponent
of the miss probability  is given by
\begin{equation}  \label{eq:prop_errorexponent_periodcluster}
\tilde{K}= \frac{1}{M}K(\tilde{\Delta};M*\SNR),
\end{equation}
where  $K(\tilde{\Delta};M*\SNR)$ is the error exponent for
uniform configuration with spacing $\tilde{\Delta}$ and $M*\SNR$ for
each sensor.
\end{proposition}

{\em Proof:} See \cite{Sung&Tong&Poor:04SPsub}.

\vspace{0.3em} The optimal detector for  periodic clustering
consists of two steps.  We first take an average of the
observations within each cluster, and then apply the optimal
 detector for a single sample at each location to the
ensemble of $N$ average values.  Intuitively, it is reasonable to
 average  the observations within a cluster since the signal
component is in the same direction and the noise is random. By
averaging, the magnitude of the signal component increases by $M$
times with the increase in the
 noise power by the same factor; the
SNR within a cluster increases by the factor $M$.  This is  shown
in the relation (\ref{eq:prop_errorexponent_periodcluster}) to
uniform configuration. The error exponent
(\ref{eq:prop_errorexponent_periodcluster}) shows the advantage
and disadvantage of the periodic clustering over uniform configuration
covering the same signal field.
  Clustering gives two benefits. First, the correlation between clusters is reduced for the same $A$
 by making
the spacing larger, and the error decay per cluster increases.
  Second, the SNR for each cluster also increases by $M$ times.  However, the effective number
  of signal samples
  is also reduced by a factor of  $M$ comparing with the uniform configuration. The performance of clustering is determined
  by the dominating factor depending on
  the diffusion rate of the underlying signal and the SNR of the observations.
\begin{figure}[htbp]
\centerline{ \SetLabels
\L(0.27*-0.1) (a) \\
\L(0.75*-0.1) (b) \\
\endSetLabels
\leavevmode
\strut\AffixLabels{
\scalefig{0.23}\epsfbox{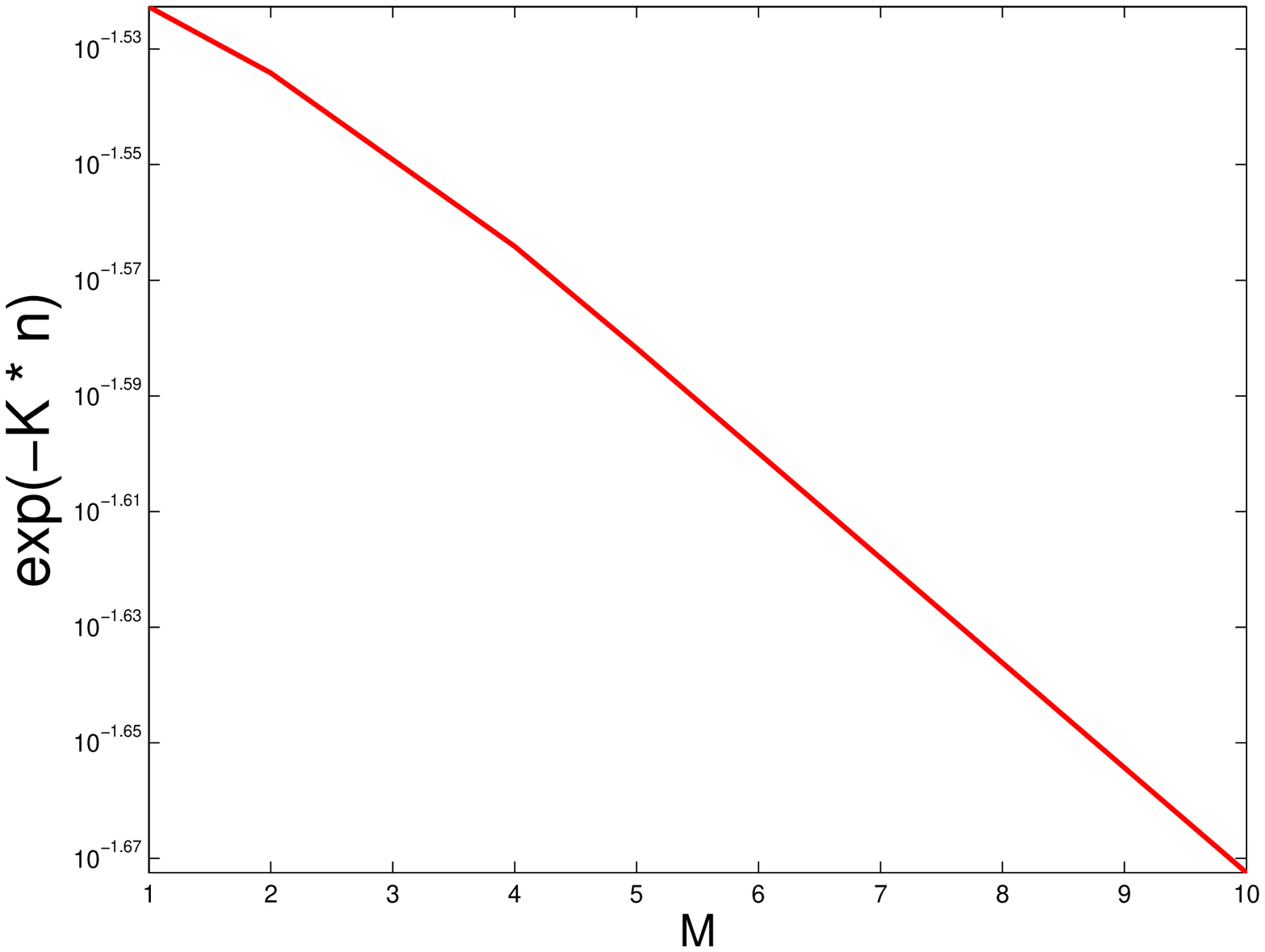}
\scalefig{0.226}\epsfbox{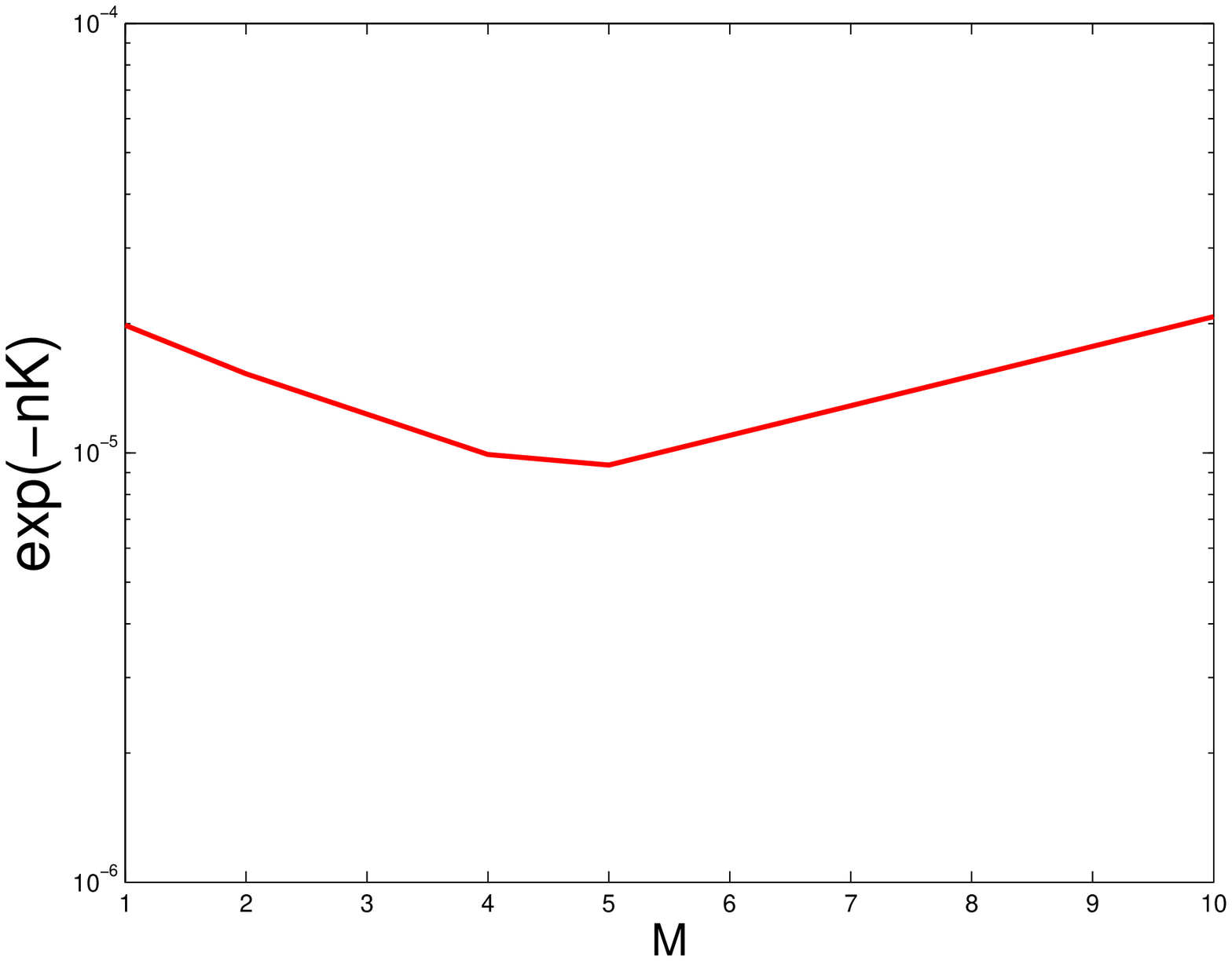}
} }
\vspace{0.3cm}
\centerline{ \SetLabels
\L(0.5*-0.1) (c) \\
\endSetLabels
\leavevmode
\strut\AffixLabels{
\scalefig{0.23}\epsfbox{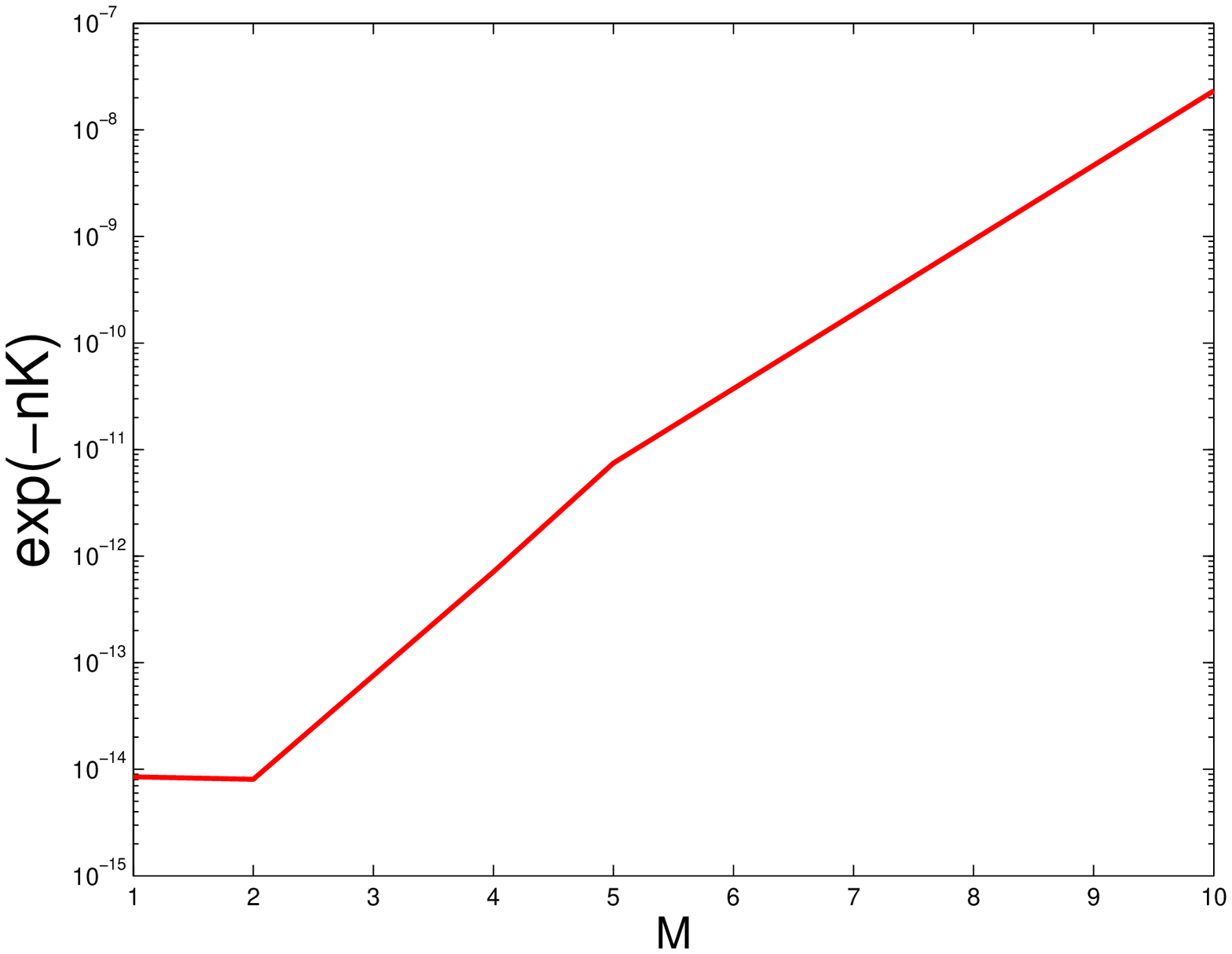}
} }
 \vspace{0.1cm} \caption{$\exp(-n\tilde{K})$ vs. $M$
($|\Sc|=1$, SNR = 10 dB): (a) $A=0.1$ (b) $A=1$ (c) $A=10$}
\label{fig:tildeKnperiodclusterhighSNR}
\end{figure}
Fig. \ref{fig:tildeKnperiodclusterhighSNR} shows
$\exp(-n\tilde{K})$, which
  is an approximate miss probability for large samples, for different diffusion rates at 10 dB SNR.
  The total number
  of sensors is fixed at $n=100$, and the cluster size is chosen as $M=[1,2,4,5,10]$.
  For the highly correlated field ($A=0.1$), it is seen that the reduced correlation between sampled signals is dominant
  and the periodic clustering gives better performance than uniform configuration (i.e., $M=1$).
  (See  Fig. \ref{fig:Kvsa}. The gain in the error exponent due to
reducing correlations is large in the high correlation region.)
  On the other hand, for the almost independent signal field ($A=10$),  clustering
    gives  worse performance than uniform configuration.
    In this case, the correlation between the scheduled samples is already weak, and  the increase
    in the error exponent due to increased spacing is insignificant as shown in Fig. \ref{fig:Kvsa}. Hence,
    the benefit of clustering results only from the increase
    in SNR.  By Theorem \ref{theo:etavsSNR}, however, the rate of increase in the error exponent due
    to the increased SNR
    is $\frac{1}{2}\log M$ at high SNR, which does not compensate for the loss in the number of effective samples
    by the factor $1/M$.
 For the signal field with intermediate correlation, there is a trade-off between the gain and loss of clustering,
 and there exists an optimal clustering as shown in Fig.\ref{fig:tildeKnperiodclusterhighSNR} (b).
 Hence, one should choose the optimal clustering depending on the diffusion rate and
 the size of the underlying
 signal field.
\begin{figure}[htbp]
\centerline{ \SetLabels
\L(0.27*-0.1) (a) \\
\L(0.75*-0.1) (b) \\
\endSetLabels
\leavevmode
\strut\AffixLabels{
\scalefig{0.23}\epsfbox{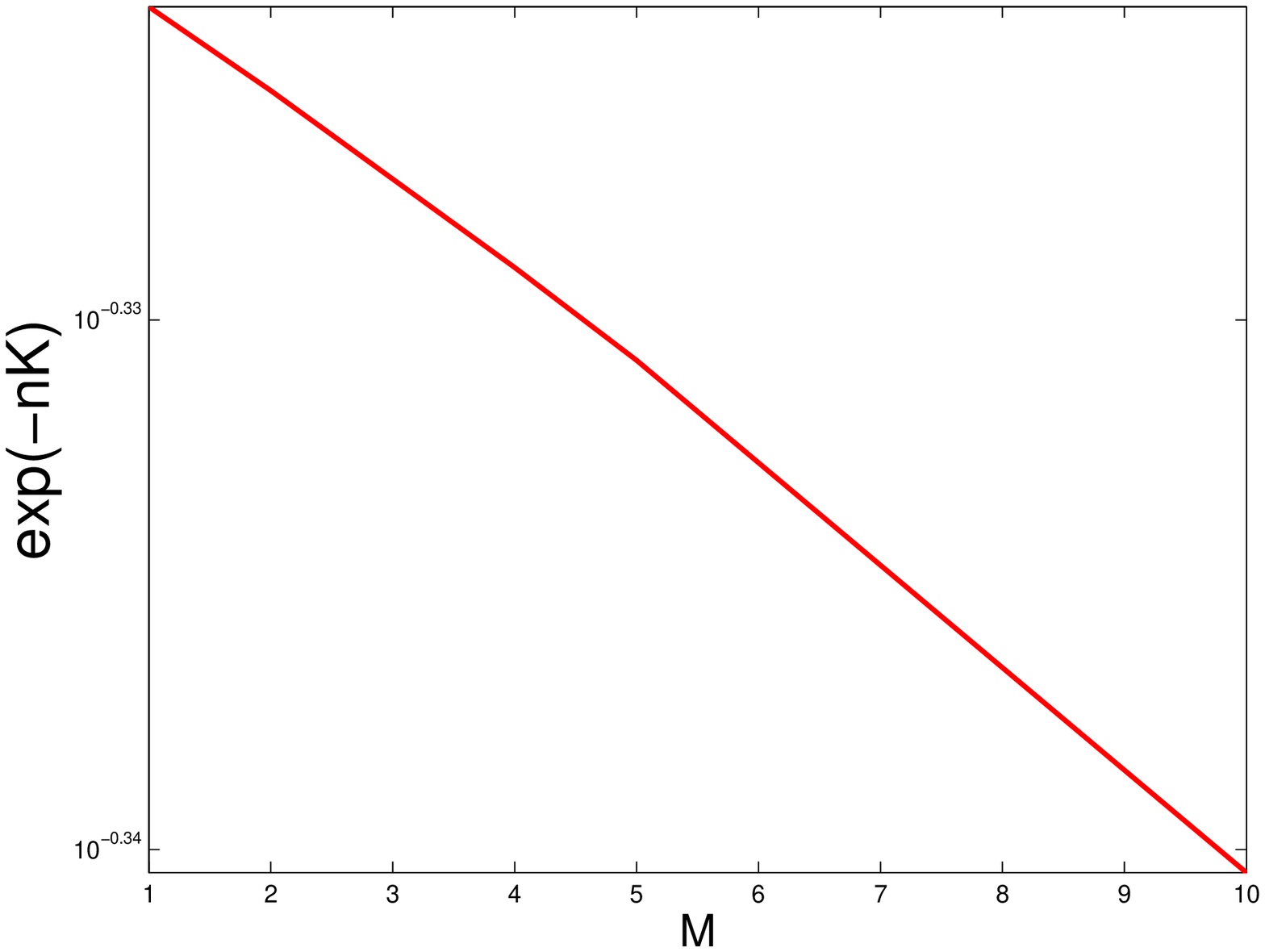}
\scalefig{0.23}\epsfbox{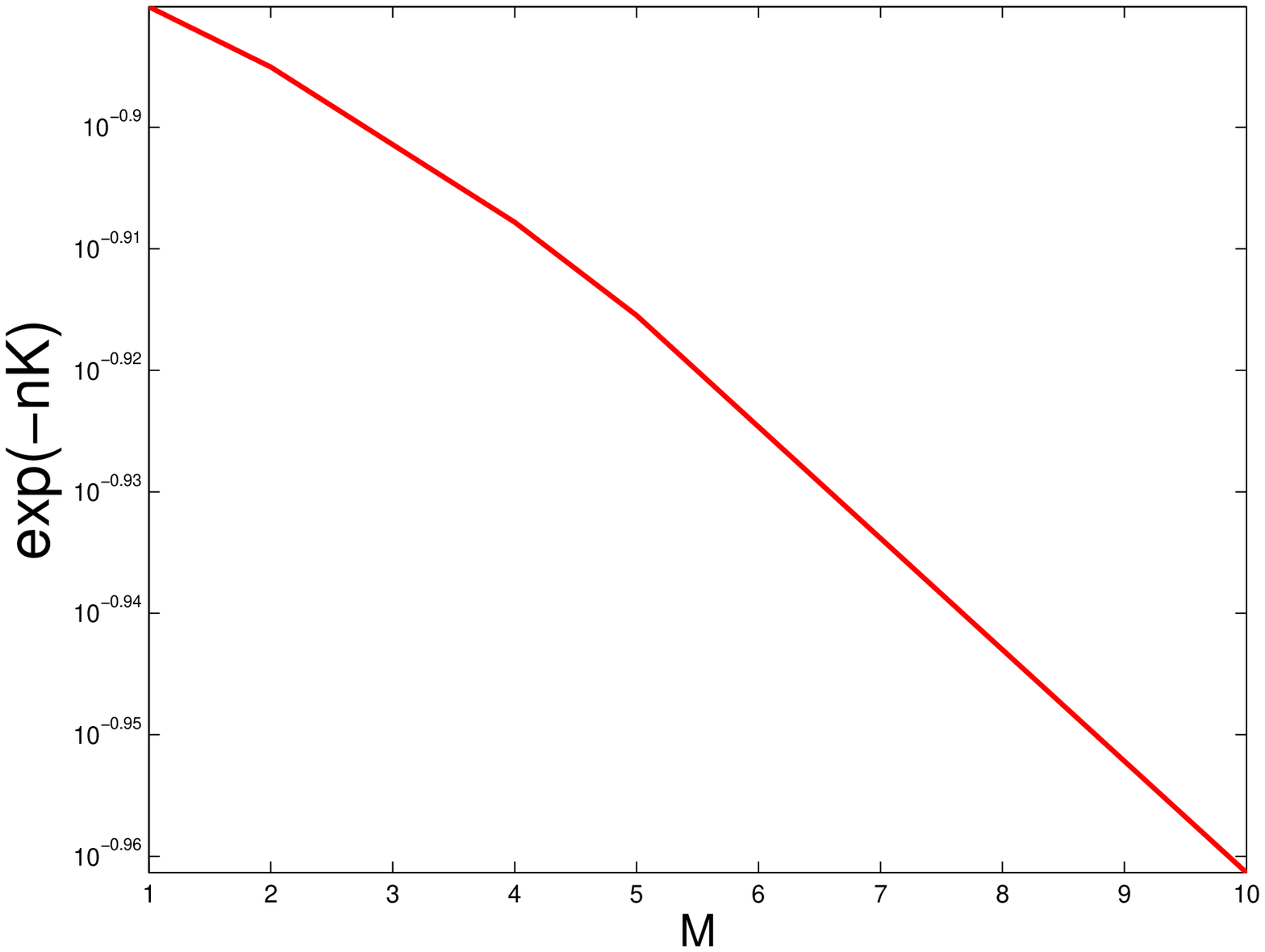}
} }
\vspace{0.3cm}
\centerline{ \SetLabels
\L(0.5*-0.1) (c) \\
\endSetLabels
\leavevmode
\strut\AffixLabels{
\scalefig{0.23}\epsfbox{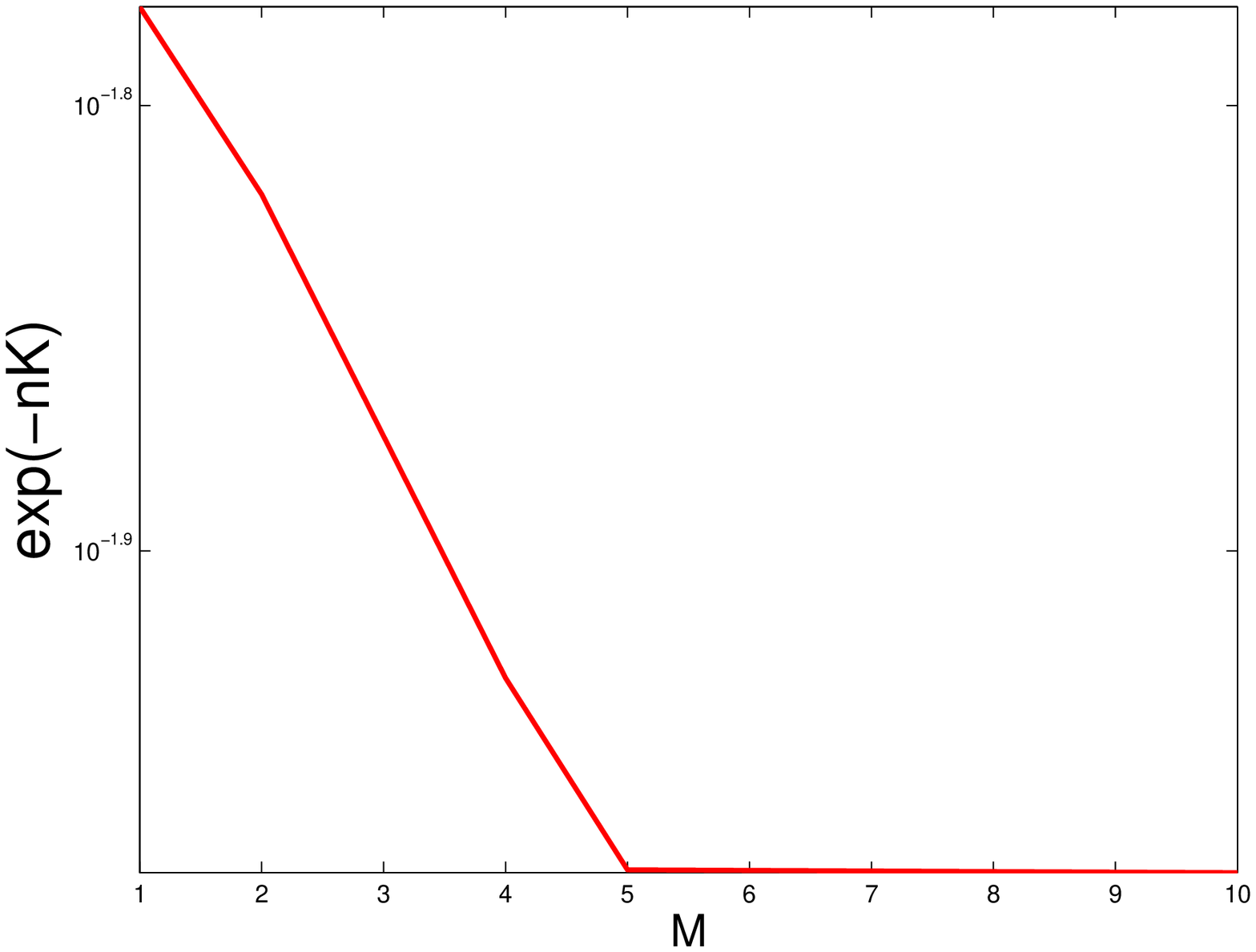}
} }
 \vspace{0.1cm} \caption{$\exp(-n\tilde{K})$ vs. $M$ ($|\Xc|=1$,
SNR = -3 dB): (a) $A=0.1$ (b) $A=1$ (c) $A=10$}
\label{fig:tildeKnperiodclusterlowSNR}
\end{figure}
\begin{comment}
\begin{figure}[htbp]
\centerline{ \SetLabels
\L(0.16*-0.1) (a) \\
\L(0.50*-0.1) (b) \\
\L(0.83*-0.1) (c) \\
\endSetLabels
\leavevmode
\strut\AffixLabels{
\scalefig{0.16}\epsfbox{figures/plot_nov4periodclusterSNRm3dBA0p1.eps}
\scalefig{0.16}\epsfbox{figures/plot_nov4periodclusterSNRm3dBA1.eps}
\scalefig{0.16}\epsfbox{figures/plot_nov4periodclusterSNRm3dBA10.eps}
} } \vspace{0.5cm} \caption{$\exp(-n\tilde{K})$ vs. $M$ ($|\Xc|=1$,
SNR = -3 dB): (a) $A=0.1$ (b) $A=1$ (c) $A=10$}
\label{fig:tildeKnperiodclusterlowSNR}
\end{figure}
\end{comment}
Fig. \ref{fig:tildeKnperiodclusterlowSNR} shows the approximate
miss probability of the periodic clustering for -3 dB SNR with
other parameters the same as in the high SNR case.  It is seen
that periodic clustering outperforms than uniform configuration in all
considered correlation values since the SNR is the dominant factor
in the detector performance at low SNR.

Clustering also explains the polynomial behavior in asymptotic
error decay for the infinite density model considered in
(\ref{eq:sensorplaceinfinitydensity}).  We can view an increase in
the number $n$ of sensors in a finite signal field as increasing
the cluster size $M$ with a fixed number $N$ of clusters. As $n$
increases, SNR per cluster increases linearly with $n$, and will
be in the high SNR regime eventually.  At high SNR, the error
exponent increases at the rate of $\frac{1}{2}\log \SNR$ by
Theorem \ref{theo:etavsSNR}.
 Hence, the overall miss probability is given approximately by
\begin{equation}
P_M \approx C \exp( - N K) = C_1 \exp( - \frac{1}{2}N \log n )\sim
C_2 n^{-1/2},
\end{equation}
for sufficiently large $n$, which coincides the results in Theorem
\ref{theo:extremecor} {\em (ii)}.  Now it is clear that for highly
correlated cases the decay in error probability with an increasing
number of sensors is mainly due to the noise averaging effect
rather than the effect of the new information about the signal in
the observations.

\subsection{Arbitrary Periodic Configuration}

The previous section shows that periodic clustering outperforms
uniform configuration depending on the field correlation. However,
periodic clustering is limited since all the sensors within a
spatial period are scheduled on the same location. Considering
periodic structure we now generalize the locations of the
scheduled sensors within a period.  First, we provide the
closed-form error exponent for the Neyman-Pearson detector for
stationary vector Gauss-Markov signals using noisy sensors.  Using
the closed-form error exponent, we investigate the optimal
periodic configuration with arbitrary locations within a period.

We again consider $n=MN$ sensors scheduled over $\Sc$ with $M$
sensors within a period, and denote the relative distance of $M$
sensors within a period as
\begin{equation}
x_1=0, ~x_2-x_1=\Delta_1, ~x_3-x_2 = \Delta_2, ~\cdots,
x_{M+1}-x_M =\Delta_M. \nonumber
\end{equation}
Hence, the interval of a period is $\Delta =
\Delta_1+\cdots+\Delta_M$. Define the signal sample and observation
vectors for period $i$ as
\begin{eqnarray}
\vec{s}_i &\defeq& [s_{1i},s_{2i},\cdots, s_{Mi}]^T, ~~i=1,\cdots, N, \label{eq:brevesbfi}\\
\vec{y}_i &\defeq& [y_{1i},y_{2i},\cdots, y_{Mi}]^T,
\end{eqnarray}
where $s_{mi} =   s_{(i-1)M+m}$ and $y_{mi}=y_{(i-1)M+m}$. The
hypotheses (\ref{eq:hypothesisscalar}) can be rewritten in vector
form as
\begin{equation}  \label{eq:hypothesisvector}
\begin{array}{lcl}
H_0 &: & \vec{y}_i = \vec{w}_i, ~~~~~~~~~~~~i=1,2,\cdots, N, \\
H_1 &: & \vec{y}_i = \vec{s}_i+ \vec{w}_i,\\
\end{array}
\end{equation}
where the measurement noise $\vec{w}_i \sim \Nc({\mathbf 0},
\sigma^2\Ibf_M)$ independent over $i$, and $\vec{s}_i$ satisfies
 the vector state-space model
\begin{eqnarray} \label{eq:statespacevector}
\vec{s}_{i+1}&=&\Abf \vec{s}_i + \Bbf \vec{u}_i, \\
\vec{u}_i &\stackrel{i.i.d.}{\sim}& \Nc(0, ~\Qbf), ~~~\Qbf \ge
0,\nn
\end{eqnarray}
where $\vec{u}_i$ are defined similarly to the quantities in
(\ref{eq:brevesbfi}). Specifically, the  feedback and input
matrices,  $\Abf$ and $\Bbf$, are given from the scalar
state-space model as {\scriptsize
\begin{equation} \label{eq:periodpatternABQ}
\Abf=
 \left[ \begin{array}{cccc}
0 & 0 & 0 & e^{-A\Delta_M}\\
0 & 0 & 0 & e^{-A(\Delta_M+\Delta_1)}\\
0 & 0 & 0 &\vdots\\
0 & 0 & 0 &e^{-A(\Delta_M+\Delta_1+\cdots+\Delta_{M-1})}
\end{array}\right],
\end{equation}
and
\begin{equation}
\Bbf= \left[ \begin{array}{cccc}
1 & 0 & 0 & 0   \\
e^{-A\Delta_1} & 1 & 0 & 0\\
\vdots & \cdots & \ddots & 0   \\
e^{-A(\Delta_1+\cdots+\Delta_{M-1})} &  & e^{-A\Delta_{M-1}} & 1
\end{array}\right],
\end{equation}
} and {\scriptsize
\begin{equation}
 \Qbf=\Pi_0
\mbox{diag}((1-e^{-2A\Delta_M}),(1-e^{-2A\Delta_1}),\cdots,(1-e^{-2A\Delta_{M-1}})).
\nonumber
\end{equation}
} Notice that $\Abf$, $\Bbf$, and $\Qbf$ are not varying with $i$
due to the periodicity. Only the last column of $\Abf$ is non-zero
due to the Markov property of the scalar process $\{s_i\}$, and
the corresponding non-zero eigenvalue of $\Abf$ is simply given by
$\lambda = e^{-A\Delta}$ so that
 $|\lambda| < 1$ for arbitrary sensor locations within a period for any diffusion rate $A > 0$.  Notice that the
 eigenvalue is the same as the correlation coefficient $a$ with sampling distance $\Delta$, the period
 of the interval.
The initial condition for the vector model is given by
\begin{equation}
\vec{s}_1 \sim \Nc(0, \Cbf_0),
\end{equation}
where  $\Cbf_0$ is given by{\scriptsize
\begin{equation}
 \Pi_0 \left[
\begin{array}{ccccc}
  1                        & e^{-A\Delta_{1,2}} & e^{-A\Delta_{1,3}} &  \cdots            & e^{-A\Delta_{1,M}}  \\
  e^{-A\Delta_{2,1}}         & 1                & e^{-A\Delta_{2,3}} &                    & e^{-A\Delta_{2,M}}  \\
  e^{-A\Delta_{3,1}}       & e^{-A\Delta_{3,3}} & 1                         &  e^{-A\Delta_{3,4}}  & \vdots                                \\
         \vdots            &                   &                          &       \ddots       & e^{-A\Delta_{1,M-1}}                    \\
  e^{-A\Delta_{M,1}} &\cdots  &                         &  e^{-A\Delta_{M-1,1}}&1                                      \\
\end{array}
\right], \nonumber
\end{equation}
} and $\Delta_{i,j} \defeq |x_j - x_i|$.
 The initial covariance matrix $\Cbf_0$ is derived from the scalar initial condition $s_1 \sim \Nc(0,\Pi_0)$,
 and it can be shown that $\Cbf_0$ satisfies the following Lyapunov equation
\begin{equation}
\Cbf_0 = \Abf \Cbf_0 \Abf^T + \Bbf\Qbf\Bbf^T.
\end{equation}
Thus, the vector signal sequence $\{\vec{s}_i\}$ is  a stationary
process
 although  the scalar process is not in general for the arbitrary periodic configuration.

For the vector case, the innovations
approach \cite{Sung&Tong&Poor:04ITsub} to obtain the error exponent
 is very  useful, and
provides a closed-form formula for the error exponent of the
Neyman-Pearson detection of stationary vector processes in noisy
observations.

\vspace{0.5em}
\begin{proposition}[Arbitrary Periodic Configuration] \label{theo:errorexponentNPvector}
For the Neyman-Pearson detector for the hypotheses
(\ref{eq:hypothesisvector},\ref{eq:statespacevector}) with level
$\alpha \in (0,1)$ (i.e. $ P_F \le \alpha$), the best error
exponent of the miss probability (per a vector observation) is
given by
\begin{equation} \label{eq:errorexponentNPtheovector}
K_v = -\frac{1}{2}\log \frac{\sigma^{2m}}{det(\Rbf_{e})}+
\frac{1}{2}\mbox{tr}\left(\Rbf_{e}^{-1}
\tilde{\Rbf}_{e}\right)-\frac{m}{2},
\end{equation}
 independently of the value of $\alpha$. The steady-state
  covariance matrices $\Rbf_{e}$
 and $\tilde{\Rbf}_{e}$ of the innovation process  calculated under $H_1$ and $H_0$, respectively,
 are given by
\begin{equation}
\Rbf_{e} =\sigma^2\Ibf_m + \Pbf,
\end{equation}
 where $\Pbf$ is the  unique stabilizing
solution of  the discrete-time Riccati equation
\begin{equation}
\label{eq:RiccatiVector} \Pbf = \Abf \Pbf \Abf^T + \Bbf\Qbf\Bbf^T -
\Abf \Pbf \Rbf_e^{-1} \Pbf \Abf^T,
\end{equation}
 and
 \begin{equation}
  \tilde{\Rbf}_{e} =
\sigma^2(\Ibf_m+ \tilde{\Pbf}),
\end{equation}
 where
$\tilde{\Pbf}$ is the unique positive-semidefinite solution of the
following Lyapunov equation
\begin{equation} \label{eq:theo1Lyapunov}
\tilde{\Pbf}= (\Abf - \Kbf_p) \tilde{\Pbf}(\Abf-\Kbf_p)^T + \Kbf_p
\Kbf_p^T,
\end{equation}
 and $\Kbf_p =
\Abf\Pbf \Rbf_{e}^{-1}$.
\end{proposition}

\vspace{0.2em}
 Proposition \ref{theo:errorexponentNPvector} is
proved by extending the results in \cite{Sung&Tong&Poor:04ITsub}
with modification from scalar to vector observations.

\begin{figure}[htbp]
\centerline{ \SetLabels
\L(0.27*-0.1) (a) \\
\L(0.75*-0.1) (b) \\
\endSetLabels
\leavevmode
\strut\AffixLabels{
\scalefig{0.24}\epsfbox{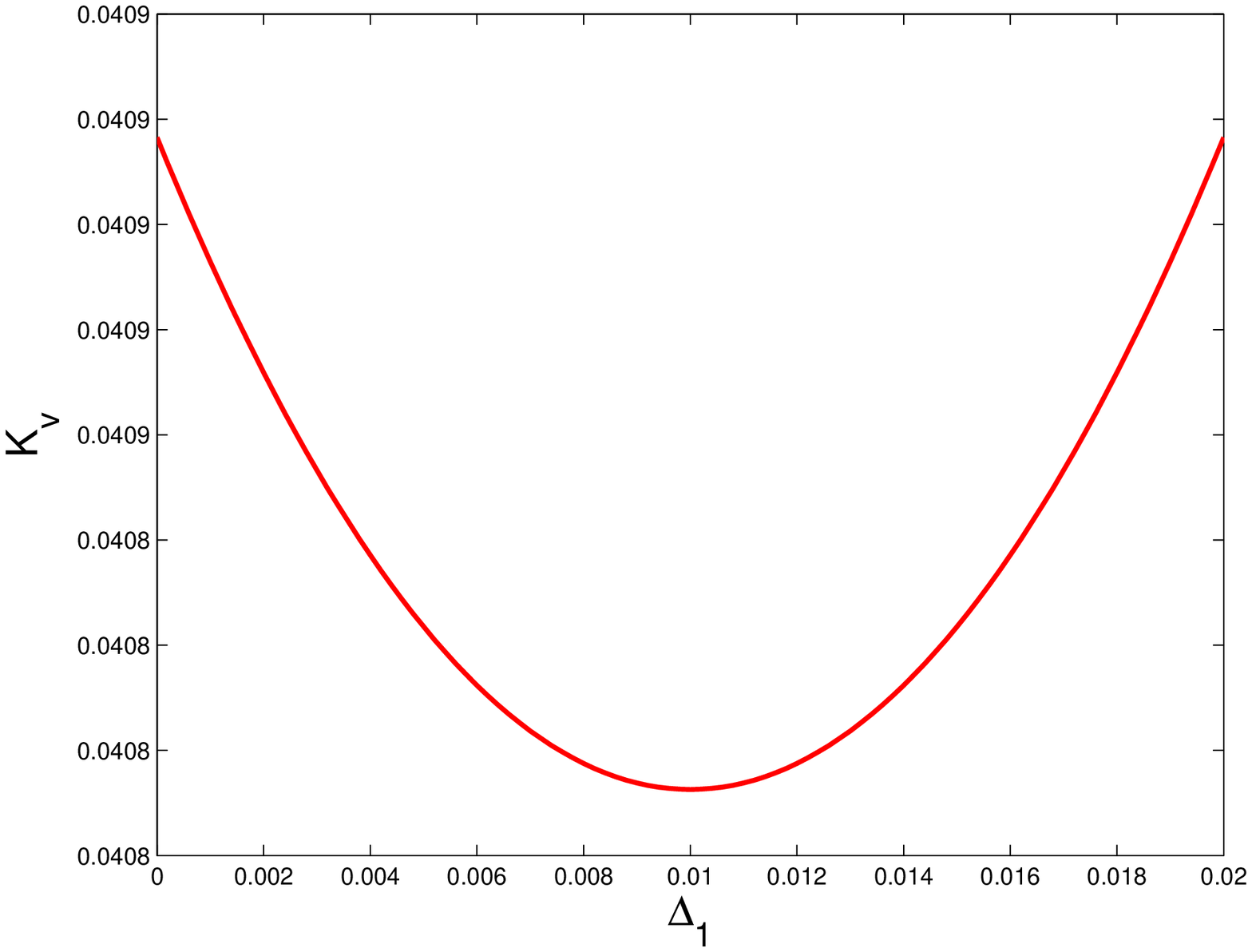}
\scalefig{0.235}\epsfbox{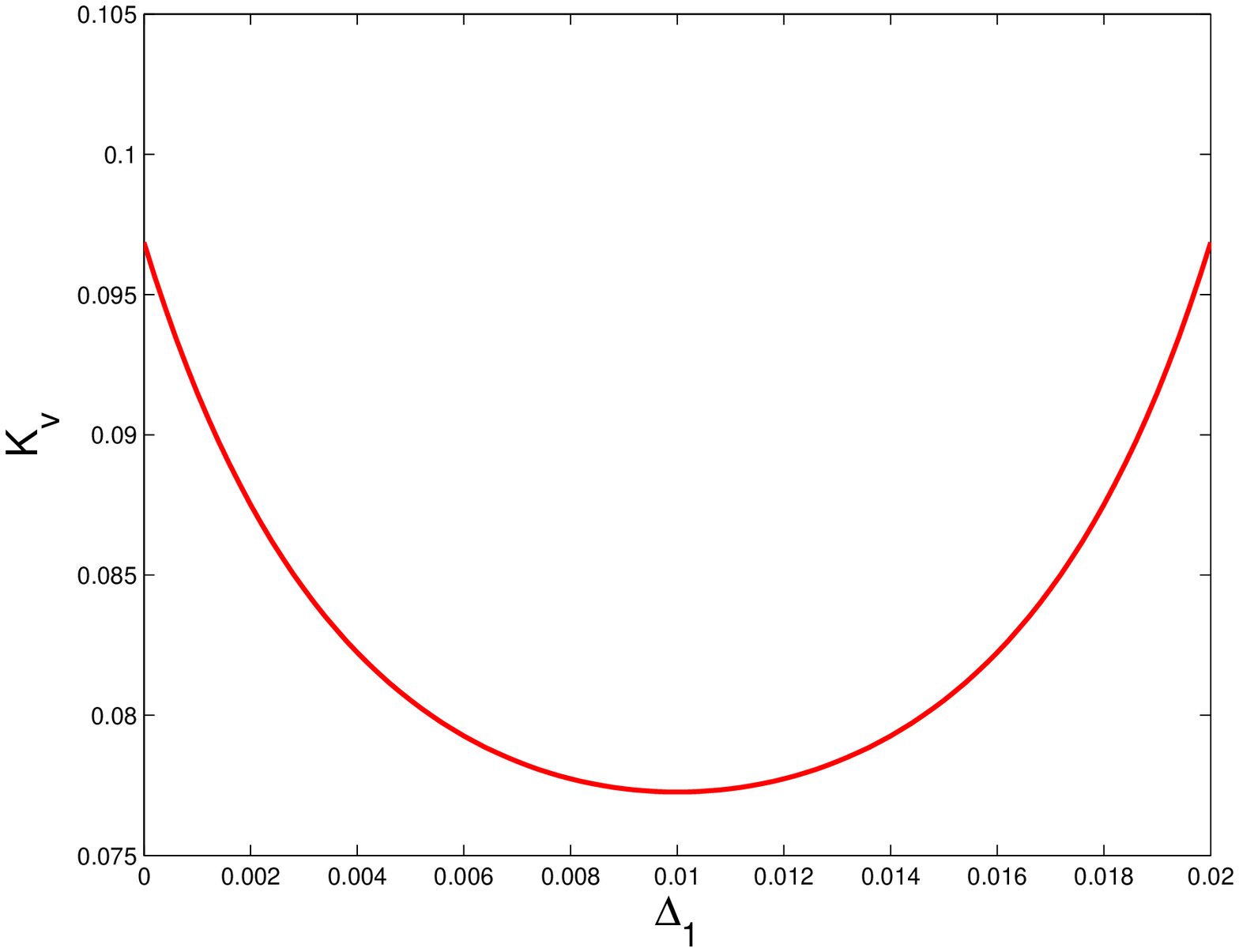}
} }
\vspace{0.3cm}
\centerline{ \SetLabels
\L(0.5*-0.1) (c) \\
\endSetLabels
\leavevmode
\strut\AffixLabels{
\scalefig{0.24}\epsfbox{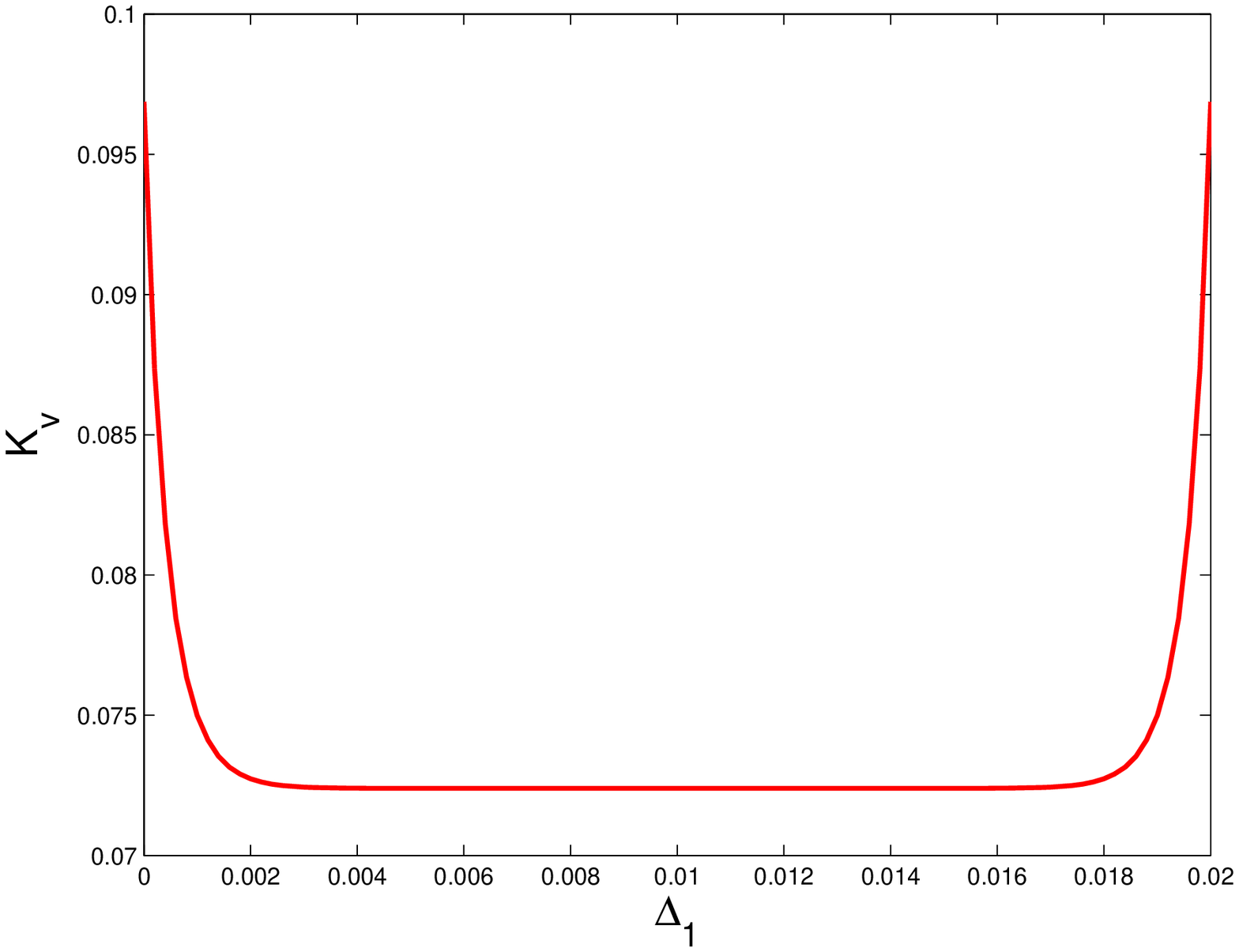}
} }
 \vspace{0.1cm} \caption{$K_v$ vs. $\Delta_1$ ($M=2$,
$\Delta=0.02$, SNR = -3 dB): (a) $A=1$ (b) $A=100$ (c) $A=1000$}
\label{fig:periodpatternM2lowSNR}
\end{figure}

Using (\ref{eq:errorexponentNPtheovector}), we now investigate the
large sample detection performance for arbitrary periodic
configuration. First, we consider the case of $M=2$ in which we have
freedom to schedule one intermediate sensor at an arbitrary
location within an interval $\Delta$. Periodic clustering and
uniform configuration are special cases of this configuration with
$\Delta_1=0$ and $\Delta_1=\Delta/2$, respectively.
\begin{figure}[htbp]
\centerline{ \SetLabels
\L(0.25*-0.1) (a) \\
\L(0.76*-0.1) (b) \\
\endSetLabels
\leavevmode
\strut\AffixLabels{
\scalefig{0.23}\epsfbox{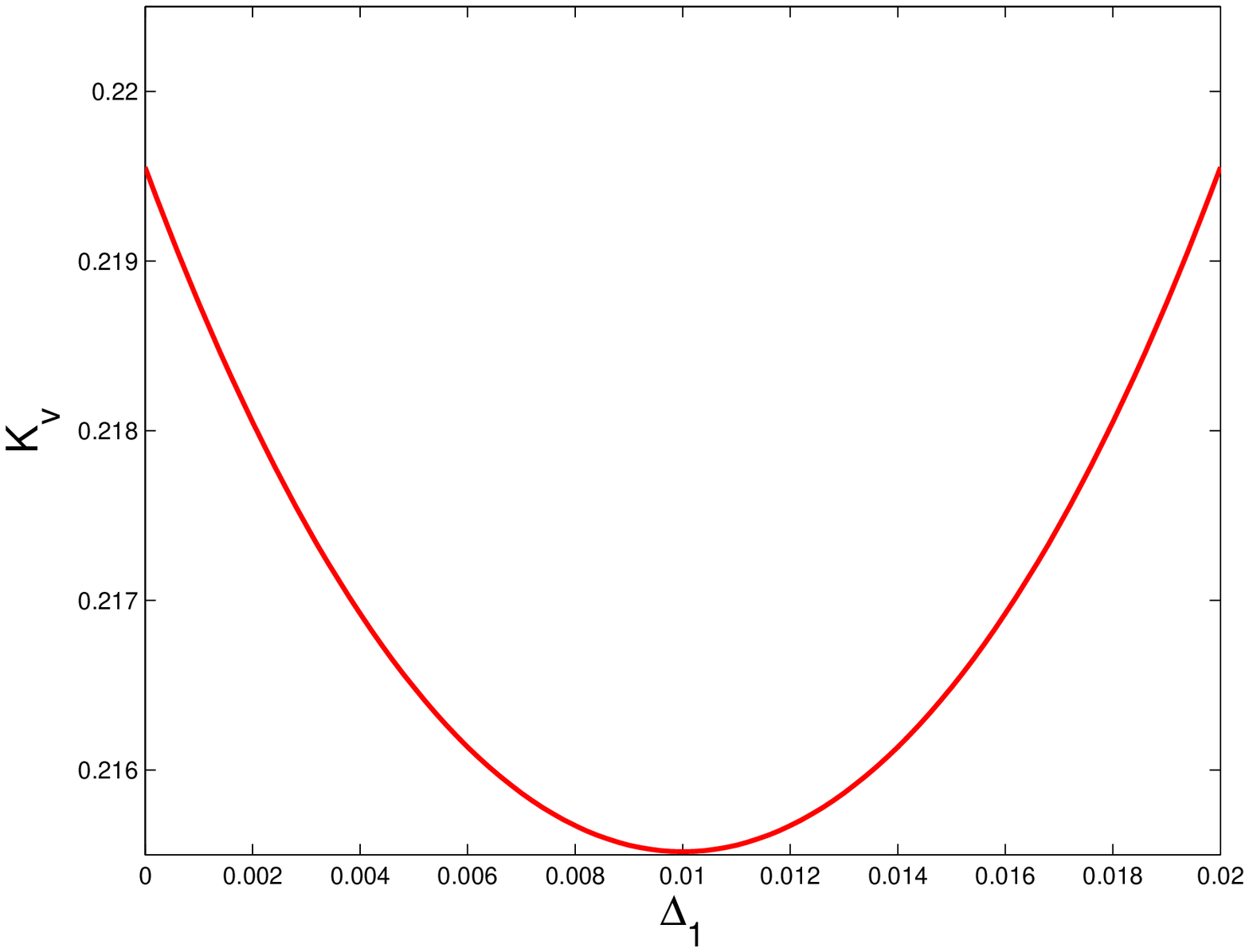}
\scalefig{0.23}\epsfbox{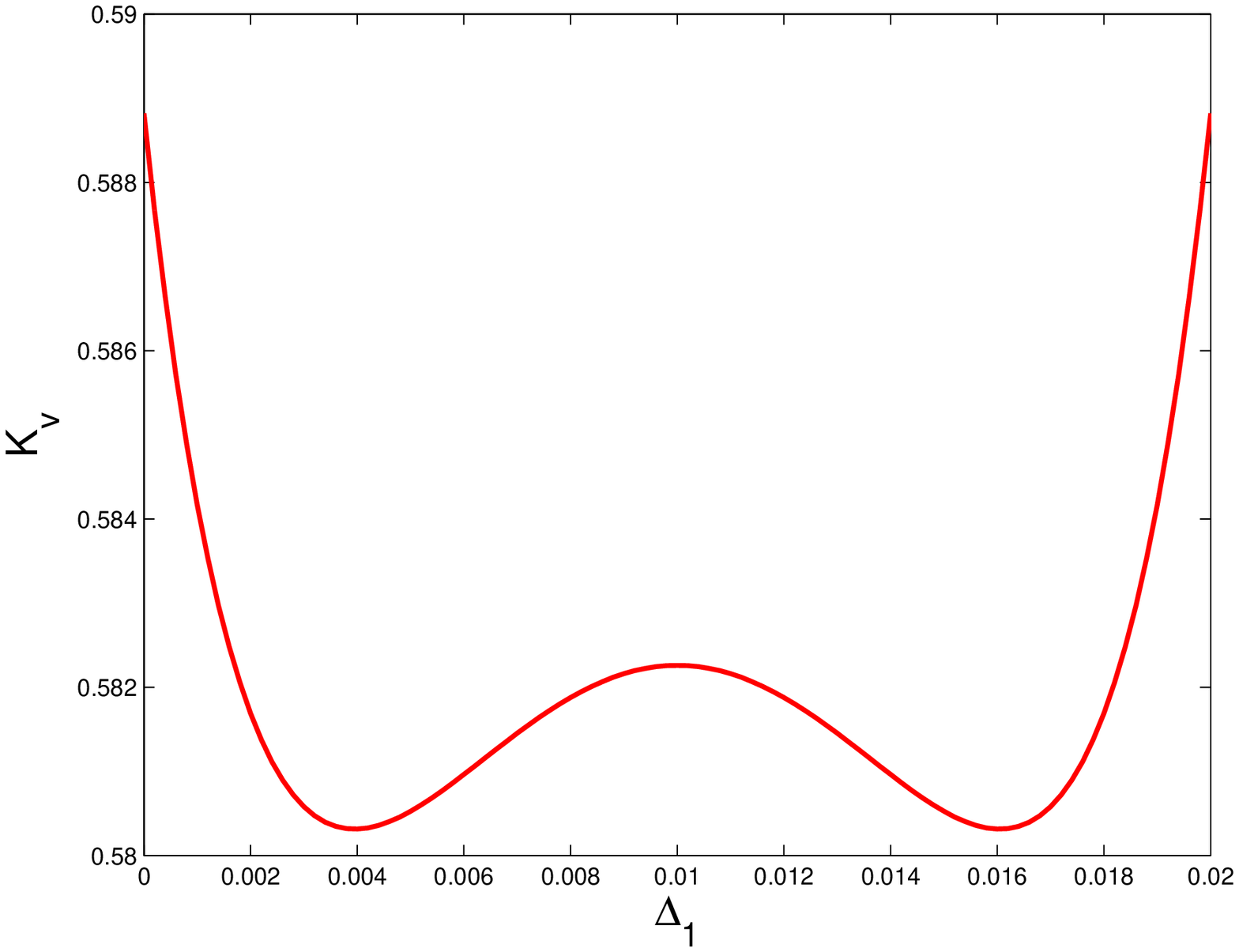}
} } \vspace{0.3cm} \centerline{ \SetLabels
\L(0.25*-0.1) (c) \\
\L(0.76*-0.1) (d) \\
\endSetLabels
\leavevmode
\strut\AffixLabels{
\scalefig{0.23}\epsfbox{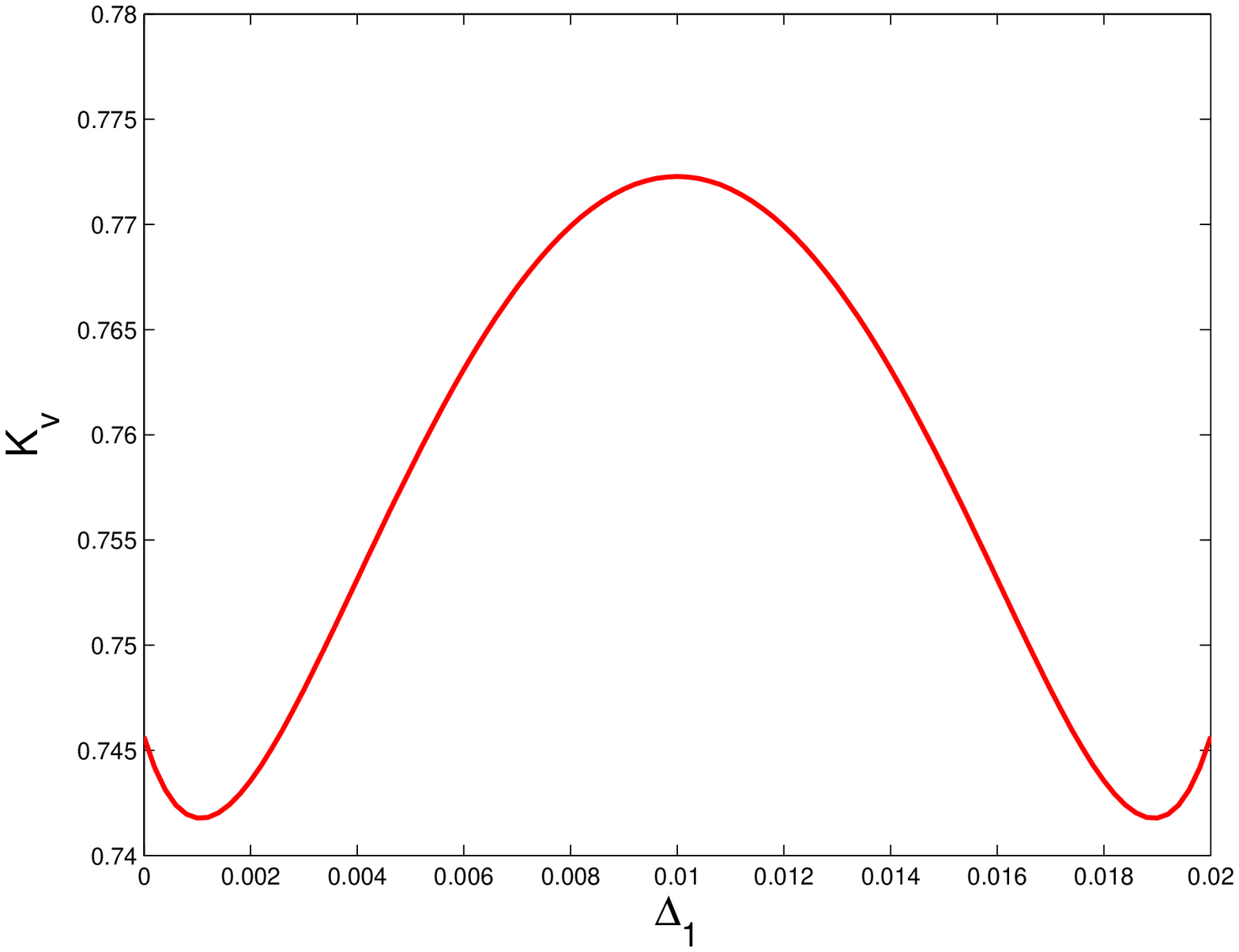}
\scalefig{0.23}\epsfbox{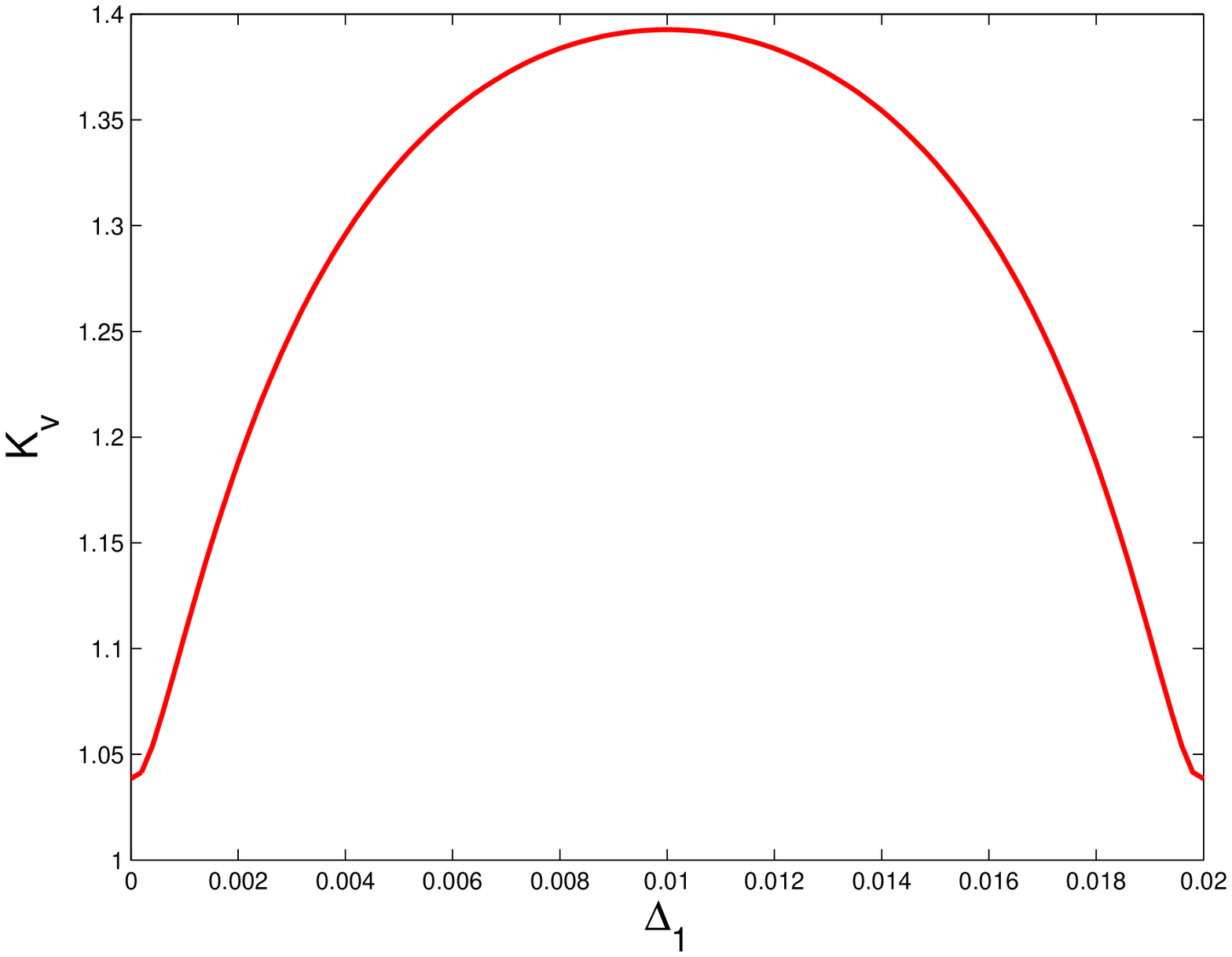}
} } \vspace{0.5cm} \caption{$K_v$ vs. $\Delta_1$ ($M=2$,
$\Delta=0.02$, SNR = 10 dB): (a) $A=1$ (b) $A=8$ (c) $A=15$ (d)
$A=100$} \label{fig:periodpatternM2highSNR}
\end{figure}
 Fig.
\ref{fig:periodpatternM2highSNR} shows the error exponent for
different diffusion rates at high SNR.  We observe an 
interesting behavior w.r.t. the diffusion rate. For the highly
correlated field ($A=1$),  periodic clustering ($\Delta_1=0$) gives the
best performance while  uniform configuration provides the worst.
However, as the field correlation becomes weak ($A=8$), we observe
a second lobe grow at the uniform configuration point ($\Delta_1 =
\Delta/2$).  The value  of the second lobe becomes larger than
that corresponding clustering as the correlation becomes weaker
($A=15$), and eventually the error exponent decreases
monotonically as the configuration deviates from  uniform
configuration to periodic clustering.  This behavior of the error exponent
clearly shows that the optimal configuration depends on the field
correlation. Consistent with the results in the previous sections,
one should reduce the correlation between observations for highly
correlated fields while the uniform configuration is best for almost
independent signal fields.  Interestingly, the optimal configuration
 for $M=2$ at high SNR is either the clustering or the uniform
configuration depending on the field correlation; no configuration
in-between is optimal!  Fig. \ref{fig:periodpatternM2lowSNR} shows
the error exponent for $M=2$ at low SNR. It is seen that
 clustering is always the best strategy for all values of field correlation
considered since the increase in the effective SNR due to noise
averaging is the dominant factor in detection performance at low
SNR. It is also seen that the location of the intermediate sensor
is not important for the highly uncorrelated field ($A=1000$)
unless it is very close to the first sensor within a period. This
is intuitively obvious since the intermediate sensor provides an
almost independent observation (for which the location does not
matter) as it separates from the first and the noise averaging is
not available between the independent samples.
\begin{comment}
\begin{figure}[htbp]
\centerline{ \SetLabels
\L(0.16*-0.1) (a) \\
\L(0.50*-0.1) (b) \\
\L(0.83*-0.1) (c) \\
\endSetLabels
\leavevmode
\strut\AffixLabels{
\scalefig{0.16}\epsfbox{figures/plot_nov6periodpatterKvsD1SNRm3dBA1.eps}
\scalefig{0.16}\epsfbox{figures/plot_nov6periodpatterKvsD1SNRm3dBA100.eps}
\scalefig{0.16}\epsfbox{figures/plot_nov6periodpatterKvsD1SNRm3dBA1000.eps}
} } \vspace{0.5cm} \caption{$K_v$ vs. $\Delta_1$ ($M=2$,
$\Delta=0.02$, SNR = -3 dB): (a) $A=1$ (b) $A=100$ (c) $A=1000$}
\label{fig:periodpatternM2lowSNR}
\end{figure}
\end{comment}

\begin{figure}[htbp]
\centerline{ \SetLabels
\L(0.25*-0.1) (a) \\
\L(0.76*-0.1) (b) \\
\endSetLabels
\leavevmode
\strut\AffixLabels{
\scalefig{0.2}\epsfbox{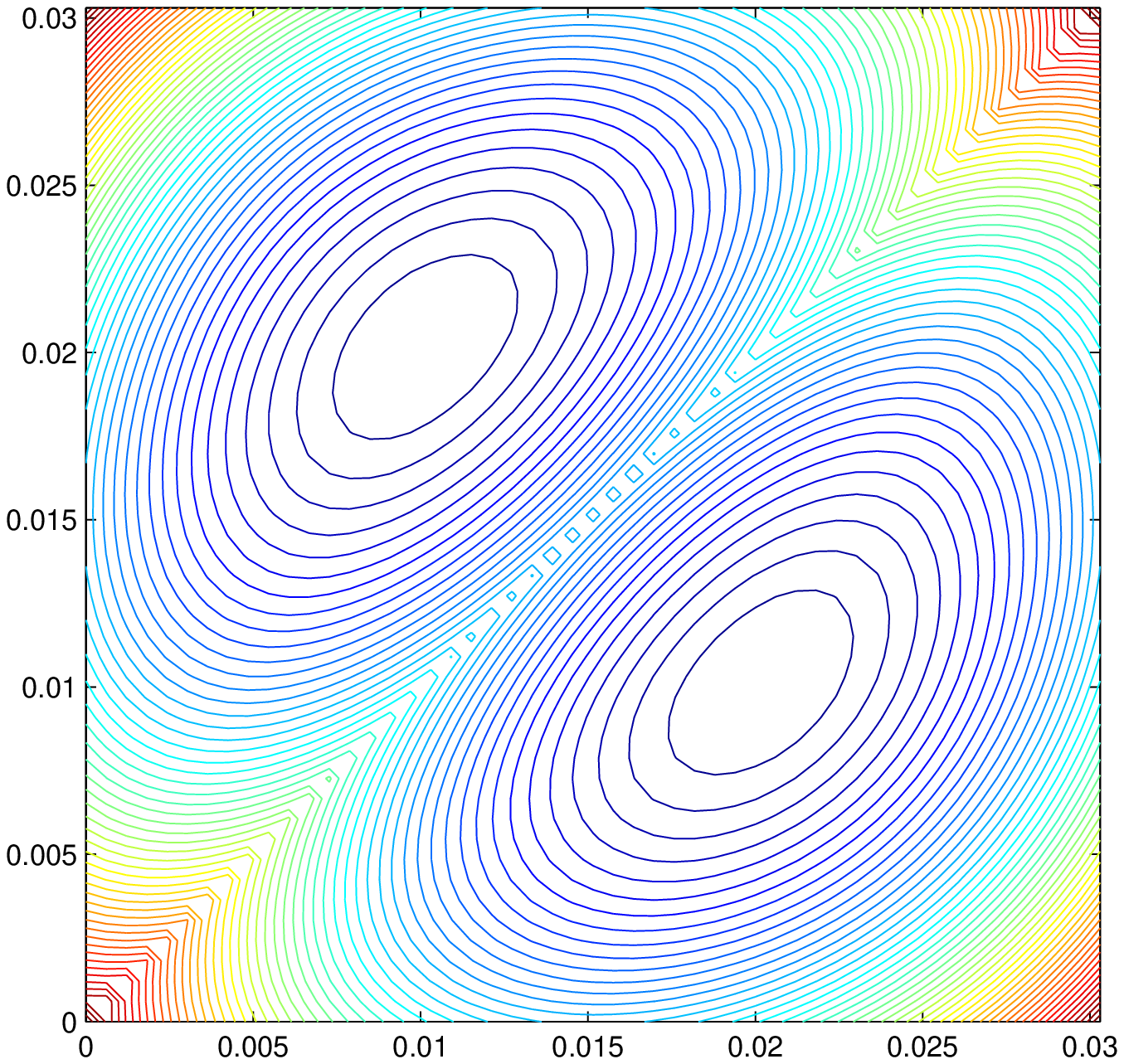}
\hspace{0.5cm}
\scalefig{0.2}\epsfbox{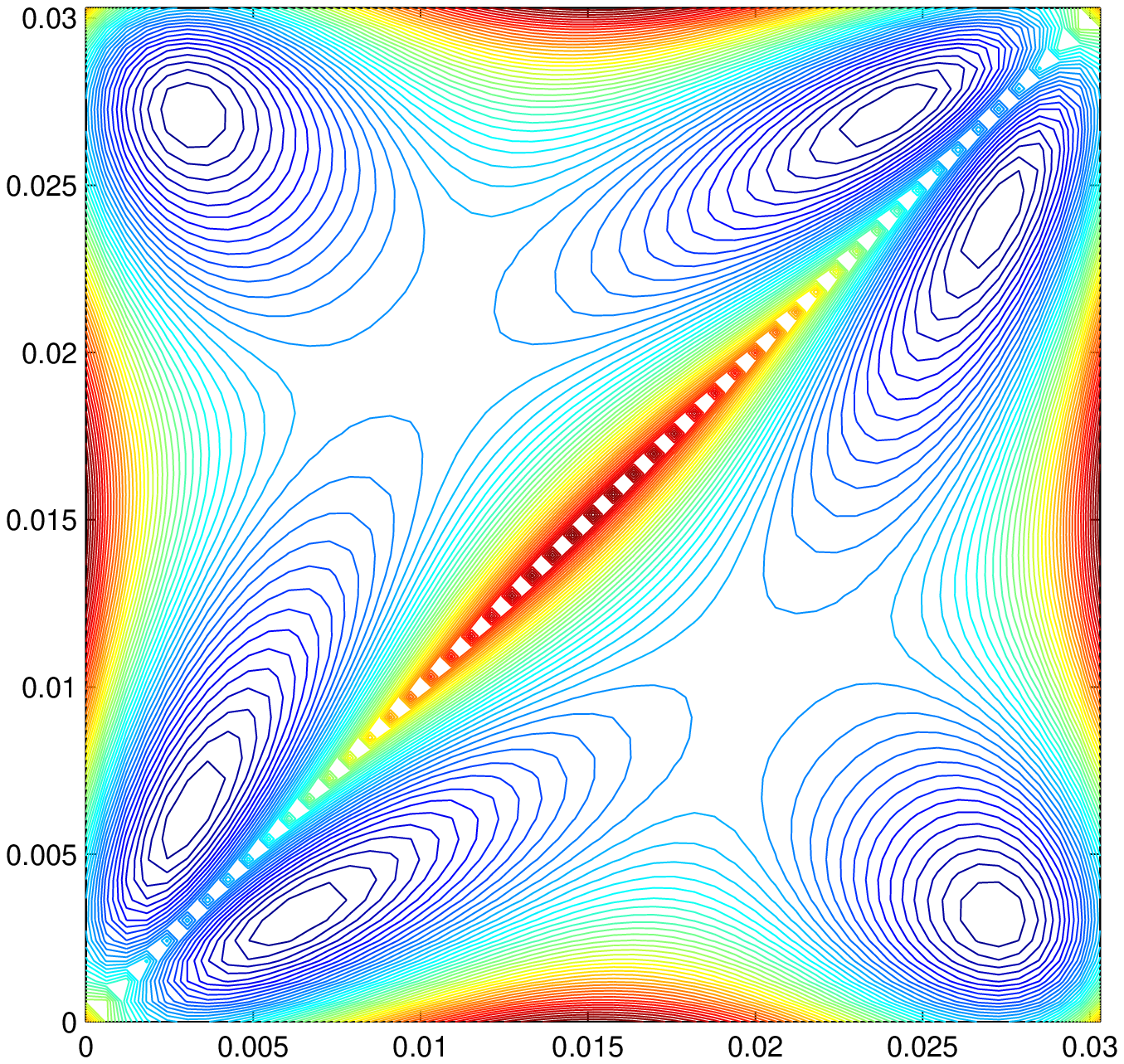}
} } \vspace{0.5cm} \centerline{ \SetLabels
\L(0.25*-0.1) (c) \\
\L(0.76*-0.1) (d) \\
\endSetLabels
\leavevmode
\strut\AffixLabels{
\scalefig{0.2}\epsfbox{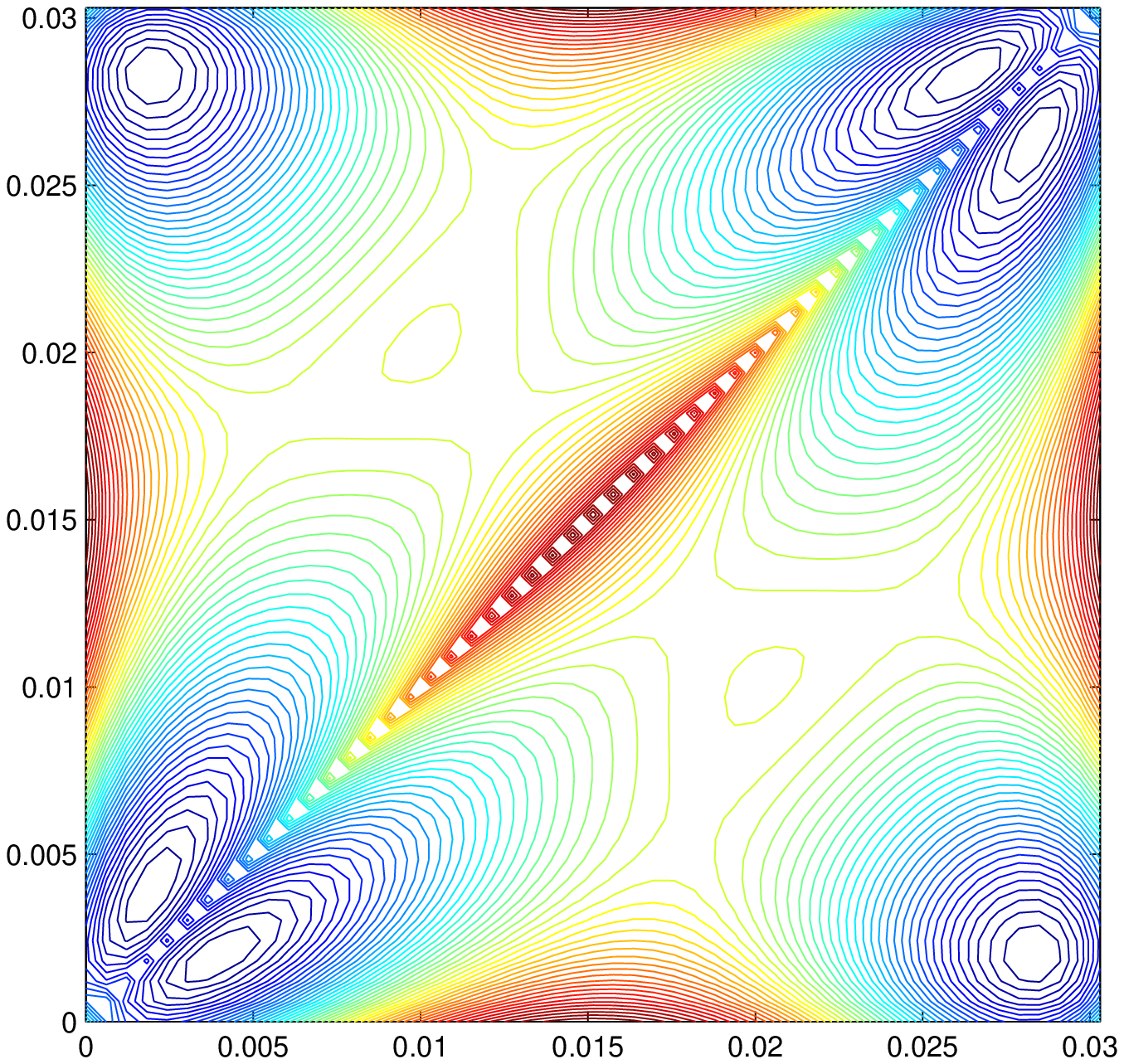}
\hspace{0.5cm}
\scalefig{0.2}\epsfbox{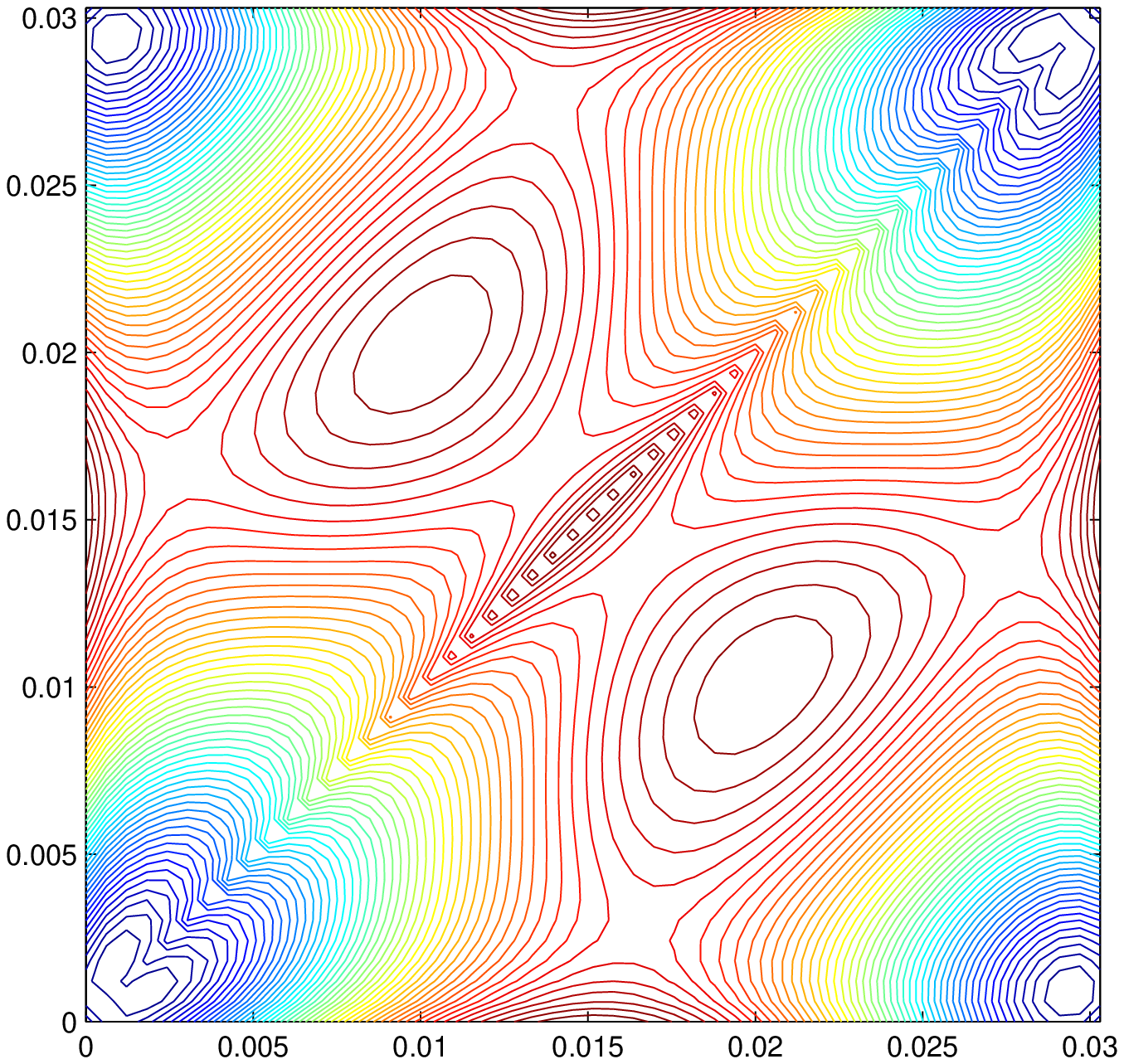}
} } \vspace{0.5cm} \caption{$K_v$ vs. $(x_2,x_3)$ ($M=3$,
$\Delta=0.03$, SNR = 10 dB): (a) $A=1$ (b) $A=5$ (c) $A=6$ (d)
$A=9$ (red:high value, blue:low value)}
\label{fig:periodpatternM3highSNR}
\end{figure}

\begin{figure}[hbtp]
\centerline{
\begin{psfrags}
 \psfrag{D1}[c]{$\Delta$}
 \psfrag{D2}[c]{$\Delta/2$}
 \psfrag{D3}[c]{$\Delta/3$}
 \psfrag{PC}[l]{{\small Periodic clustering}}
 \psfrag{Uniform}[l]{{\small Uniform configuration}}
 \psfrag{sensor}[l]{{\small Activated sensor}}
 \psfrag{FC}[l]{{\small Field correlation}}
 \psfrag{Strong}[l]{{\small Strong}}
 \psfrag{Weak}[l]{{\small Weak}}
\scalefig{0.38}\epsfbox{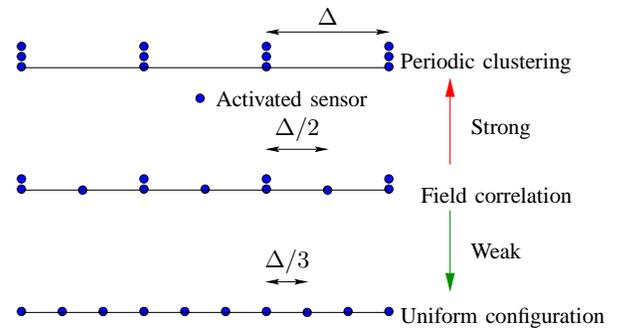}
\end{psfrags}
} \caption{Optimal configuration for $M=3$ (SNR = 10 dB)}
\label{fig:optimalconfigM3}
\end{figure}

For the case of $M=3$, due to the periodicity, the location of one
sensor in a period is fixed and we can choose the locations $(x_2,
x_3)$ of the two other sensors arbitrarily such that $ 0 \le x_2
\le \Delta$ and $0 \le x_3 \le \Delta$.  Fig.
\ref{fig:periodpatternM3highSNR} shows the error exponent  as a
function of  $(x_2,x_3)$ for $M=3$ at 10 dB SNR. Similar behavior
is seen as in the case of $M=2$.
For the highly correlated signal ($A=1$), we see the maximal value
of the error exponent at $(0,0)$,$(0,\Delta)$,$(\Delta,0)$, and
$(\Delta,\Delta)$, which all correspond to periodic clustering.
Hence, periodic clustering is the best among all configurations.
In this case, it is seen that uniform configuration is the worst
configuration.   As the field correlation becomes weak, however,
the best configuration moves to  uniform configuration eventually, as
seen in Fig \ref{fig:periodpatternM3highSNR} (d). It is seen that
placing two sensors clustered and one in the middle of the spatial
period is optimal for transitory values of field correlation as
shown in Fig. \ref{fig:periodpatternM3highSNR} (b) and (c). Figure \ref{fig:optimalconfigM3} summarizes the optimal configuration for different field correlation.
Other
results confirm that periodic clustering gives the best
configuration for most values of field correlations at low SNR.

\subsection{Sensor Placement}

So far we have assumed that sensors have already been deployed and
considered the activation for sensing and transmissions from the selected
sensors.  A related problem is sensor placement.  When $n$ sensors
are planned to be deployed over a signal field for the detection
application, how should we place the $n$ sensors in the field? The
results in the previous sections provide the answer for this
problem as well.

\section{Conclusion}
\label{sec:conclusion}

We have considered  energy-efficient sensor activation for  large sensor
networks deployed to detect  correlated random fields.
 Using our  results on large-sample error behavior in this application,
we have analyzed and compared the detection capabilities of
different sensor configuration strategies.  The optimal configuration
is a function of the field correlation and the SNR of sensor
observations.  For uniform configuration, the scheduled sensors
should be maximally separated to cover the entire signal field for
SNR $>1$. For SNR $<1$, on the other hand, there exists an optimal
spacing between the scheduled sensors.  We have also derived the
error exponents of periodic clustering and arbitrary periodic
configuration. Periodic clustering may outperform  uniform configuration 
depending on the field correlation  and SNR. Furthermore, there
exists an optimal cluster size for intermediate values of
correlation.  The closed-form error exponent obtained for the
vector state-space model explains the transitory error behavior
from  periodic clustering to uniform configuration.


{\scriptsize
\bibliographystyle{plain}

}

\end{document}